\documentclass[12pt]{article}
\usepackage{amsmath}
\usepackage{amsfonts}        
\usepackage{amssymb}
\usepackage{amsbsy}
\usepackage{mathrsfs}
\usepackage{epsfig}      
\usepackage{color}       

\usepackage[T1]{fontenc} 

\usepackage{graphicx}

\newlength{\TZ}
\DeclareFontFamily{OT1}{pzc}{}
\DeclareFontShape{OT1}{pzc}{m}{it}{<-> s * [1.200] pzcmi7t}{}
\DeclareMathAlphabet{\mathpzc}{OT1}{pzc}{m}{it}
\newcommand{\BEQ}{\begin{equation}}     
\newcommand{\BEA}{\begin{eqnarray}}
\newcommand{\BD}{\begin{displaymath}}
\newcommand{\EEQ}{\end{equation}}       
\newcommand{\EEA}{\end{eqnarray}}
\newcommand{\ED}{\end{displaymath}}

\newcommand{\vep}{\varepsilon}          
\newcommand{\D}{{\rm d}}                
\newcommand{\II}{{\rm i}}               
\newcommand{\demi}{\frac{1}{2}}         
\newcommand{\wit}[1]{\widetilde{#1}}    
\newcommand{\wht}[1]{\widehat{#1}}      
\newcommand{\lap}[1]{\overline{#1}}     
\newcommand{\fR}{\mathfrak{R}}          
\newcommand{\bR}{\lap{\mathfrak{R}}}    

\renewcommand{\vec}[1]{\boldsymbol{#1}} 
\newcommand{\fns}{\footnotesize}        

\newcommand{\appsection}[2]{\setcounter{equation}{0}\setcounter{subsection}{0}
\section*{Appendix #1. #2}
\renewcommand{\theequation}{#1.\arabic{equation}}
              \renewcommand{\thesection}{#1}
              \renewcommand{\thefigure}{#1\arabic{figure}}\setcounter{figure}{0} }

\catcode`\@=11
\def\numberbysection{\@addtoreset{equation}{section}
        \def\theequation{\thesection.\arabic{equation}}}
\numberbysection                        

\definecolor{gruen}{rgb}{0,0.625,0}       
\definecolor{rot}{rgb}{0.75,0,0}          
\definecolor{blau}{rgb}{0,0,0.75}         
\definecolor{casta}{rgb}{0.45,0.20,0}     
\definecolor{gelb}{rgb}{0.825,0.725,0.0}  

\newcommand{\BLAU}[1]{\textcolor{black}{{\rm #1}}}	

\begin{document}

\begin{titlepage}

\vskip 1.5 cm
\begin{center}
{\LARGE \bf Correlators in phase-ordering from Schr\"odinger-invariance}
\end{center}

\vskip 2.0 cm
\centerline{{\bf Malte Henkel}$^{a,b}$ and {\bf Stoimen Stoimenov}$^c$}
\vskip 0.5 cm
\centerline{$^a$Laboratoire de Physique et Chimie Th\'eoriques (CNRS UMR 7019),}
\centerline{Universit\'e de Lorraine Nancy, B.P. 70239, F -- 54506 Vand{\oe}uvre l\`es Nancy Cedex, France}
\vspace{0.5cm}
\centerline{$^b$Centro de F\'{i}sica Te\'{o}rica e Computacional, Universidade de Lisboa,}
\centerline{Campo Grande, P -- 1749-016 Lisboa, Portugal}
\vspace{0.5cm}
\centerline{$^c$ Institute of Nuclear Research and Nuclear Energy, Bulgarian Academy of Sciences,}
\centerline{72 Tsarigradsko chaussee, Blvd., BG -- 1784 Sofia, Bulgaria}
\vspace{0.5cm}

\begin{abstract}
Systems undergoing phase-ordering kinetics after a quench into the ordered phase with $0<T<T_c$ from 
a fully disordered initial state and with a non-conserved order-parameter have the dynamical exponent $\mathpzc{z}=2$.
The long-time behaviour of their single-time and two-time correlators, determined by the noisy initial conditions, is derived from Schr\"odinger-invariance.  
It is shown that the generic ageing scaling forms of the single-time and two-time correlators follow from the Schr\"odinger-covariance of the four-point response functions.
The autocorrelation exponent $\lambda$ is related to the passage exponent $\zeta_p$ which describes
the time-scale for the cross-over into the ageing regime. Both Porod's law and the bounds $d/2 \leq \lambda \leq d$ are reproduced in a simple way.
The dynamical finite-size scaling in fully finite systems is established. 
The dynamical scaling of global correlators are found and the low-temperature generalisation $\lambda= d-2\Theta$ of the
Janssen-Schaub-Schmittmann scaling relation is derived.
\end{abstract}
\end{titlepage}

\setcounter{footnote}{0}

\section{Introduction} \label{sec:1}

\begin{figure}[tb]  
\includegraphics[width=.98\hsize]{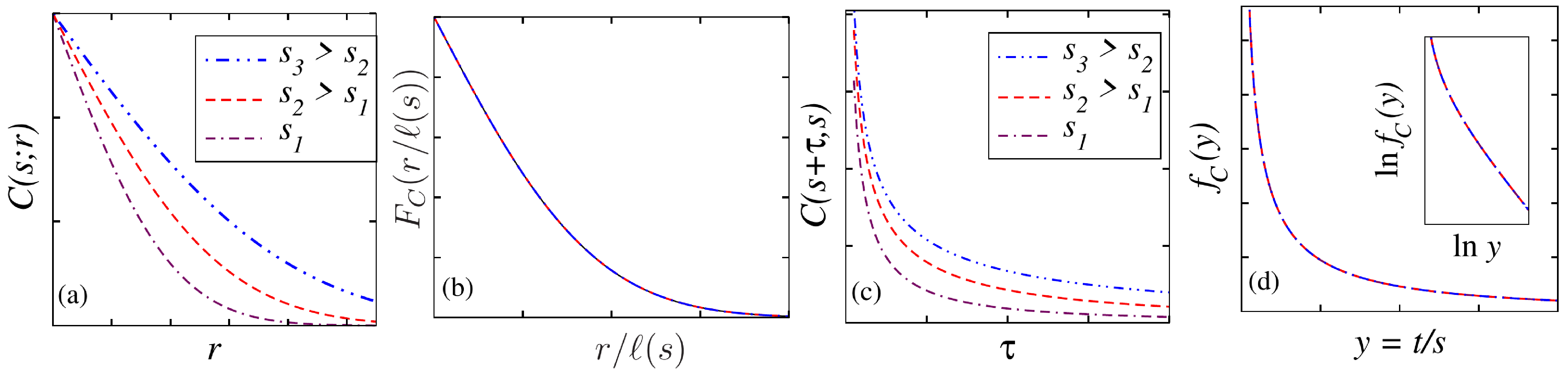}  
\caption[fig1]{Schematic illustration of physical ageing in single-time and two-time correlators.
A typical single-time correlator $C(s;r)$ is in panel (a) for different times
$s_1<s_2<s_3$. In panel (b), their collapse when replotted over against rescaled lengths $r/\ell(s)$
is shown where $\ell(s)\sim s^{1/\mathpzc{z}}$ is the dynamical length scale.
In panel (c) a typical two-time auto-correlator $C(s+\tau,s)$ is displayed over against $\tau=t-s$, for different waiting times $s_1<s_2<s_3$. These data
collapse when replotted in panel (d) over against $y=t/s$. The inset shows the asymptotic power-law form $f_C(y)\sim y^{-\lambda_C/z}$.
\label{fig1} }
\end{figure}

Phase-ordering kinetics \cite{Bray94a}, after its pioneering beginnings in the 1960s \cite{Lifs61,Lifs62,Wagn61}, 
is an important sub-class of more generic physical ageing behaviour \cite{Stru78,Cugl03,Cugl15,Vinc24}. 
It is usually brought about by preparing a classical system initially in a
totally disordered state before quenching it to a temperature $T<T_c$, into the ordered phase which occurs below the equilibrium critical temperature $T_c>0$.
Due to the competition of at least two distinct, but equivalent, equilibrium states, 
such systems \BLAU{become spatially inhomogeneous and} decompose microscopically into ordered clusters of a time-dependent and growing linear size $\ell(t)$. 
Throughout, \BLAU{this growth will be assumed to occur} algebraically $\ell(t)\sim t^{1/\mathpzc{z}}$ which defines the dynamical
exponent $\mathpzc{z}$. A convenient characterisation uses the coarse-grained and time-space-dependent order-parameter $\phi(t,\vec{r})$
-- which in magnetic systems is the local magnetisation. 
In phase-ordering, with a model-A-type dynamics of the order-parameter which does not satisfy any macroscopic conservation law, 
it is well-established that $\mathpzc{z}=2$ \cite{Bray94b}. We shall consider
throughout a description in terms of continuum fields which has been used many times in the past to address this much-studied problem,
see \cite{Bray94a,Godr02,Cugl03,Maze06,Puri09,Henk10,Taeu14,Cugl15} and references therein.
Useful tools for the study of such systems include the single-time and {\em two-time correlation function} $C$ and the {\em two-time response function} $R$, defined as
\BEQ \label{gl:1}
C(t,s;\vec{r}) = \bigl\langle \phi(t,\vec{r})\phi(s,\vec{0})\bigr\rangle \;\; , \;\;
R(t,s;\vec{r}) = \left. \frac{\delta \bigl\langle \phi(t,\vec{r})\bigr\rangle}{\delta h(s,\vec{0})}\right|_{h=0}
=\bigl\langle \phi(t,\vec{r})\wit{\phi}(s,\vec{0})\bigr\rangle
\EEQ
where the average is both over initial states as well as over thermal histories. For a totally disordered initial state we have
$\left\langle \phi(t,\vec{r})\right\rangle=\left\langle \phi(0,\vec{r})\right\rangle=0$, 
and implicitly shall also admit throughout the usual spatial translation- and rotation-invariances,
such that $\vec{r}\mapsto r = |\vec{r}|$ for notational simplicity.
In (\ref{gl:1}) we also anticipate from Janssen-de Dominicis theory \cite{Domi76,Jans76} that $R$ can be formally rewritten as a correlator with the so-called
{\em response scaling operator} $\wit{\phi}(t,\vec{r})$.
Set $t=s$ in (\ref{gl:1}) to obtain\footnote{Practical means of obtaining the length scale $\ell(t)$ include
solving an equation $C\bigl(t;\ell(t)\bigr)=\mathfrak{c}$ with a constant $0<\mathfrak{c}<1$, or calculating the second moment
$\ell^2(t) = \left.\int_{\mathbb{R}^d}\!\D\vec{r}\: r^2\, C(t;{r})\right/\int_{\mathbb{R}^d}\!\D\vec{r}\:C(t;{r})$.}
the {\em single-time correlator} $C(s;r) := C(s,s;r)$.
Setting $r=0$ in (\ref{gl:1}) produces the {\em auto-correlator} $C(t,s):=C(t,s;0)$ and the {\em auto-response} $R(t,s):=R(t,s;0)$.
The typical long-time behaviour of both $C(s;r)$ and $C(t,s)$ is shown in figure~\ref{fig1}
and serves to illustrate the three defining properties of physical ageing \cite{Stru78,Henk10}.
First, there is {\em slow relaxation dynamics} since the correlators evolve more slowly when $s$ is increasing, see figure~\ref{fig1}ac.
Second, since the correlators depend on both $s$ and $r$, or $s$ and $\tau=t-s$ respectively, there is {\em no time-translation-invariance}. Third, figure~\ref{fig1}bd shows
a data collapse occurs when the same data are replotted over against $r/\ell(t)$ or $y=t/s$, respectively, and illustrates {\em dynamical scaling}.
Formally, this can be cast into the scaling forms, \BLAU{already specific for phase-ordering kinetics}
\begin{subequations} \label{gl:local}
\BEQ \label{gl:2}
C(t,s;{r}) =          F_C\left( \frac{t}{s}; \frac{\bigl|\vec{r}\bigr|}{s^{1/2}}\right) \;\; , \;\;
R(t,s;{r}) = s^{-1-a} F_R\left( \frac{t}{s}; \frac{\bigl|\vec{r}\bigr|}{s^{1/2}}\right)
\EEQ
with the {\em ageing exponent} $a$. In addition, one usually finds asymptotically for $y\gg 1$
\BEQ \label{gl:3a}
f_C(y) = F_C(y,0) \sim y^{-\lambda_C/2} \;\; , \;\; f_R(y) = F_R(y,0) \sim y^{-\lambda_R/2}
\EEQ
where $\lambda_C$ is the {\em auto-correlation exponent} \cite{Huse89} and $\lambda_R$ is the {\em auto-response exponent} \cite{Pico02}. Also, it is generically seen that
\BLAU{(at least for spatially short-ranged initial correlations \cite{Bray91a})}
\BEQ \label{lambda}
\lambda = \lambda_C = \lambda_R
\EEQ
\end{subequations}
The following bounds are known: $d/2\leq \lambda\leq d$  \cite{Fish88a,Yeun96a}.
The scaling functions $F_{C,R}\bigl(y,r s^{-1/2}\bigr)$ in (\ref{gl:2}) are expected {\em universal}, that is their form 
should be independent of microscopic `details', such as the lattice structure, the precise form of the interactions, or
temperature. {If this can be described by a continuum limit, 
there are renormalisation group arguments \cite{Bray90,Bray94a} that the existence and the properties of the scaling functions
should be described by a $T=0$ fixed point}.\footnote{The continuum limit admitted here and the $T\to 0$
limit in lattice models do {\bf not} always commute. We explicitly exclude
from our considerations the rich and peculiar features which can arise after a quench exactly to zero temperature $T=0$ an a lattice, e.g. in the $3D$ Ising model
\cite{Corb08a,Olej11a,Olej11b,Chris21,Agra21,Agra22,Gess23,Gess24}.} For a scalar order-parameter, the single-time correlator $C(s;{r})$ typically has a cusp
at $r=0$, see figure~\ref{fig1}b, which is known in the literature as {\em Porod's law} \cite{Poro51}.

Rather than focussing on the study of specific models, or on analytic theories based on uncontrolled approximations,
we shall inquire into the dynamical scaling (\ref{gl:2}) and shall ask to what extent this scaling behaviour and \BLAU{especially} if the  form of the universal functions $F_{C,R}$
appearing therein can be determined from dynamical symmetries of phase-ordering alone.
Since at least for model-A-type dynamics without any conservation law of the order-parameter $\phi(t,\vec{r})$,
it is well-known that $\mathpzc{z}=2$ \cite{Bray94b}, it looks promising to study the extension of the naturally realised dynamical scaling to the full group of
{\em Schr\"odinger-transformations}. \BLAU{These are by definition the transformations of time and space \cite{Hage72,Nied72,Jack72}
\BEQ
t \mapsto t' = \frac{\alpha t+\beta}{\gamma t + \delta} \;\; , \;\; 
\vec{r} \mapsto \vec{r}' = \frac{\mathscr{R} \vec{r} + \vec{v} t + \vec{a}}{\gamma t +\delta} \;\; ; \;\; \alpha\delta -\beta\gamma =1
\EEQ
where $\mathscr{R}\in\mbox{\sl SO}(d)$ is a rotation matrix and $\vec{v}, \vec{a}\in\mathbb{R}^d$ are vectors. 
These transformations were already studied in the 19$^{\rm th}$ century by Jacobi and Lie;} for a historical review, see \cite{Duva24}.
\BLAU{Which physical observables can one expect to transform covariantly under Schr\"odinger-transformations~? 
\begin{enumerate}
\item At equilibrium, systems at a critical point are scale-invariant \cite{Card96} and under certain conditions are also thought to be conformally invariant, according to a well-known
argument involving the symmetry and tracelessness of the energy-momentum tensor \cite{Call70}. But a precise reasoning is considerably more subtle \cite{Polc88}; 
see \cite{Riva05,Fort11,Fort12,Dela16,Naka15,Naka20,Gime24} for various aspects of this on-going and controversial discussion. 
In conformally-invariant theories, multi-point correlators $\bigl\langle \phi \cdots \phi\bigr\rangle$ 
built from `quasi-primary' or `primary' scaling operators $\phi$ are expected to transform covariantly. 
Notably for two space dimensions, this conformal covariance can be used to determine their precise functional form \cite{BPZ,Fran97}. 
Conformal bootstrap techniques are currently being explored to generalise this to $d>2$ dimensions \cite{Pola19,Rych23}. 
It is a long-standing objective to try to extend conformal invariance to dynamics \cite{Card85}, especially in situations far from equilibrium. 
\item Non-equilibrium field-theory, described through Janssen-de Dominicis theory \cite{Domi76,Jans76}, requires besides the order-parameter $\phi$ 
the introduction of a conjugate response operator $\wit{\phi}$. 
It arises from the coupling of the system with external reservoirs, which may include heat baths or noisy initial conditions \cite{Jans92,Taeu14}. 
Studying the properties of $\wit{\phi}$ will become essential below. 
\item Far from equilibrium, correlators $C=\bigl\langle \phi\phi\bigr\rangle$ and responses $R=\bigl\langle \phi \wit{\phi}\,\bigr\rangle$ are distinct and have different properties. 
For phase-ordering kinetics, both obey dynamical scaling \cite{Bray94a} with $\mathpzc{z}=2$ in the important special case of model-A-type dynamics \cite{Bray94b}
without macroscopic conservation laws. 
\item Available evidence from models suggests that the responses $R$ in phase-ordering kinetics indeed transform covariantly 
when the dynamical scaling symmetry is extended to the Schr\"odinger algebra \cite{Henk10}.\footnote{The difficult question whether dynamical scaling, together with
Galilei-invariance, really implies full Schr\"odinger-invariance in sufficiently `local' theories, is not studied here. 
We just assume that the formal treatment \cite{Henk03a} of the energy-momentum tensor and its conservation laws is sufficient.}
In contrast with the true representations of the semi-simple conformal algebra, the Schr\"odinger Lie algebra is not semi-simple and its
representations are necessarily projective \cite{Barg54,Levy63,Simm71,Perr77,Azca95,Unte12,Hall13}. \\
~~In non-equilibrium field-theory the response operator $\wit{\phi}$ takes over the r\^ole of a `complex conjugate' of $\phi$ which mathematically is needed for the projective
representations of the Schr\"odinger algebra \cite{Barg54,Azca95,Unte12,Hall13}. 
\item Projective representations lead to Bargman super-selection rules \cite{Barg54} such that a response 
$R=\bigl\langle \phi \wit{\phi}\,\bigr\rangle$ found from Schr\"odinger-covariance can be non-trivial whereas a Schr\"odinger-covariant correlator 
$\bigl\langle \phi\phi\bigr\rangle=0$ must vanish. \\
~~Hence physical correlators $C$ cannot be found from a covariance requirement 
but must be derived by reducing them to higher multi-point (and Schr\"odinger-covariant) response functions \cite{Pico04,Henk10}.
\end{enumerate}}

\BLAU{Since the Schr\"odinger Lie algebra does contain the time translations $X_{-1}=-\partial_t$, it cannot be used directly a dynamical symmetry of a non-equilibrium system which is not
time-translation-invariant. But recently,} it has been shown that the generic forms of both two-time responses and two-time correlators far from equilibrium can be obtained by
changing the representation of the dynamical symmetry Lie algebra, which was assumed to consist only of dilatations and generalised time-translations \cite{Henk25c}. 
\BLAU{The hypothesis of covariance of both responses and correlators under this representation led to:} 
\begin{enumerate}
\item[(I)] the complete known generic phenomenology of physical ageing (\BLAU{both phase-ordering and non-equilibrium critical dynamics with generic $\mathpzc{z}$}) 
of the two-time observables with $t>s$ can be reproduced from these two symmetry requirements, in agreement with numerous specific studies from models and
\item[(II)] the non-linear terms in the equation of motion of the order-parameter $\phi(t,\vec{r})$ become irrelevant (in the renormalisation-group sense)
in the long-time limit if only the auto-correlation exponent obeys the condition $\lambda>1$.
\end{enumerate}
\BLAU{This means that the non-equilibrium behaviour can be obtained from a time-dependent change of representation of a Lie algebra which {\em a priori} can serve only
as dynamical symmetry of an equilibrium system.}\footnote{In systems which undergo non-equilibrium critical dynamics after a quench onto $T=T_c$ and which in addition have 
$\mathpzc{z}=2$, explicit expressions for the single-time correlator $C\bigl(s;{r}\bigr)$
and the two-time auto-correlator $C(t,s)$ could be derived from the covariance of the three-point response function $\langle \phi\phi\wit{\phi^2}\rangle$ \cite{Henk25d}.
The results are reproduced in the $1D$ Glauber-Ising model quenched to temperature $T=0$, the spherical model quenched onto $T=T_c>0$ and very recently also the voter model
in $d>0$ dimensions \cite{Henk25f}.} \BLAU{However, in spite of this phenomenological success, the assumption of covariance of the correlator $C$, even under such a small Lie algebra, 
should be considered as merely heuristic, since the equal-time case $t=s$ cannot be included \cite[Appendix B]{Henk25c}.} 

\BLAU{In consequence, we revert to an approach where {\em only} the response $R$ can be assumed to transform covarianty. Correlators $C$ have to be obtained from a reduction
to response functions.} For phase-ordering kinetics this means that certain four-point response functions 
$\langle \phi\phi\wit{\phi}\wit{\phi}\rangle$ must be found, see eq.~(\ref{corrFT}) in section~\ref{sec:2}. 
\BLAU{At equilibrium, these should be found via Schr\"odinger-covariance \cite{Golk14,Shim21,Volo09} and then the result should be brought} out-of-equilibrium by the above-mentioned
change of representation. \BLAU{In contrast to non-equilibrium dynamics, it is less obvious for phase-ordering kinetics if such a mapping could work, since the corresponding ordered
equilibrium state is not scale-invariant.} It is one of the objectives of this work to test this idea, specifically for the four-point functions. 
\BLAU{{}From this conceptually clean basis we can first re-derive all correlators and no longer need to distinguish artificially 
between two-time correlators (where the argument of \cite{Henk25c} applies)
and single-time correlators (where it does not).} 
We shall show that 
\begin{enumerate}
\item all standard expectations (\ref{gl:local}) on the ageing of local two-time correlators are reproduced.
\BLAU{This also includes all properties (I) mentioned above. On the other hand, we must accept being now restricted to $\mathpzc{z}=2$.} 
\item single-time correlators can be included and in particular the consistency with long-standing properties such as 
{\sc Porod}'s law \cite{Poro51,Bray94a} can be checked.
\end{enumerate} 
Extensions to finite-size scaling and global correlators follow as well and in particular we shall re-derive the extension $\lambda=d-2\Theta$ of the celebrate
critical-point {\sc Janssen-Schaub-Schmittmann} \cite{Jans89} scaling relation to all temperatures $T<T_c$ below criticality. 
\BLAU{The implication of Schr\"odinger-invariance that the dynamical exponent $\mathpzc{z}=2$ is no restriction, since that value is known to hold naturally \cite{Bray94b},
at least for model-A-type dynamics.} 
Since all this follows from Schr\"odinger-invariance alone and no model-specific information enters, the universality of these results is evident. 
\BLAU{On the other hand, we shall also find that Schr\"odinger-invariance alone is not strong enough to explicitly fix the form of the scaling functions 
$F_C$.\footnote{This is distinct from what occurs in non-equilibrium critical dynamics \cite{Henk25d}.}}

This work is organised as follows. In section~\ref{sec:2} we recall the necessary background on non-equilibrium field theory \BLAU{of phase-ordering kinetics}, 
Schr\"odinger-invariance and the new extension of its representations to far-from-equilibrium situations. 
Section~\ref{sec:3} recalls first the essential properties of the two-time response function before the
calculation of two-time and single-time correlators is carried out in detail and their generic properties are described, which includes the discussion of finite-size
effects and global two-time correlators. We conclude in section~\ref{sec:4}. Four appendices treat the calculation of four-point response functions (including systems of finite-size),
the absence of change in form of the two-point response function in the presence of dimensionful variables and {\sc Porod}'s law.

\section{Background} \label{sec:2}

The forthcoming calculation of correlators in section~\ref{sec:3} will be based on four distinct ingredients which we shall now briefly review.

\subsection{Non-equilibrium field-theory}

This work is inserted into the context of non-equilibrium continuum field-theory \cite{Domi76,Jans76,Jans92,Taeu14}.
The average of an observable $\mathscr{A}$ is found in principle from the functional integral
\BEQ \label{dynft}
\bigl\langle \mathscr{A}\bigr\rangle = \int \mathscr{D}\phi\mathscr{D}\wit{\phi}\; \mathscr{A}[\phi]\, e^{-{\cal J}[\phi,\wit{\phi}]}
\EEQ
which already includes probability conservation \cite{Taeu14}. 
For \BLAU{phase-ordering kinetics, with} non-conserved dynamics of the order-parameter (`model A'), the Janssen-de Dominicis action reads
\begin{subequations} \label{actionJdD}
\begin{align}
{\cal J}[\phi,\wit{\phi}\,] &= {\cal J}_0[\phi,\wit{\phi}\,] + {\cal J}_b[\wit{\phi}\,] \label{actionJdDa} \\
{\cal J}_0[\phi,\wit{\phi}\,] &= \int \!\D t\D\vec{r}\: \left( \wit{\phi} \left( \partial_t - \Delta_{\vec{r}} - V'[\phi]\right)\phi  \right) \\
{\cal J}_b[\wit{\phi}\,] &= -\demi \int_{\mathbb{R}^{2d}} \!\D\vec{R}\D\vec{R}'\: \wit{\phi}(0,\vec{R}) C_0\bigl(\vec{R}-\vec{R}'\bigr) \wit{\phi}(0,\vec{R}') \label{actionJdDc} 
\end{align}
\end{subequations}
with the interaction $V'[\phi]$ and the spatial laplacian $\Delta_{\vec{r}}$ (and the usual re-scalings). The noise contribution is assumed here to contain only the `initial'
correlator $C_0\bigl(\vec{R}\bigr)$, implicitly at some microscopic time which refers to the beginning of the scaling regime. 
Throughout, we shall assume $C_0(\vec{R})$ to be spatially short-ranged.
Thermal noise is considered irrelevant \cite{Bray90,Bray94a}.
The {\em deterministic action} ${\cal J}_0[\phi,\wit{\phi}\,]$ is then used to define the {\em deterministic averages} $\bigl\langle \cdot \bigr\rangle_0$ as
\BEQ \label{detave}
\bigl\langle \mathscr{A}\bigr\rangle_0 = \int \mathscr{D}\phi\mathscr{D}\wit{\phi}\; \mathscr{A}[\phi]\, e^{-{\cal J}_0[\phi,\wit{\phi}]}
\EEQ
Causality considerations \cite{Jans92,Cala05,Taeu14} or Local-Scale-Invariance ({\sc lsi}) \cite{Pico04} then imply the Barg\-man su\-per\-se\-lec\-tion rules \cite{Barg54}
\BEQ\label{Bargman}
\left\langle \overbrace{~\phi \cdots \phi~}^{\mbox{\rm ~~$n$ times~~}}
             \overbrace{ ~\wit{\phi} \cdots \wit{\phi}~}^{\mbox{\rm ~~$m$ times~~}}\right\rangle_0 \sim \delta_{n,m}
\EEQ
for the deterministic averages. Only observables built from an equal number of order-parameters $\phi$
and conjugate response operators $\wit{\phi}$ can have non-vanishing deterministic averages.
Response functions as in (\ref{gl:1}) can now be found directly as deterministic averages.
It is straightforward to see that a two-time response $R=\left\langle\phi\wit{\phi}\right\rangle=\left\langle\phi\wit{\phi}\right\rangle_0$
or a four-point response $\left\langle\phi\phi\wit{\phi}\wit{\phi}\right\rangle=\left\langle\phi\phi\wit{\phi}\wit{\phi}\right\rangle_0$
does not depend explicitly on the noise \cite{Pico04}.
\BLAU{However, any attempt to find correlators from a covariance requirement is futile, since the Bargman rules imply that $\left\langle\phi\phi\right\rangle_0=0$.} 
Therefore, correlators such as $\left\langle\phi\phi\right\rangle$ are obtained from higher-point response functions \cite{Pico04,Henk10}
\BLAU{\BEA
C(t,s;r)
&=& \bigl\langle \phi(t,\vec{r}+\vec{r}_0)\phi(s,\vec{r}_0) \bigr\rangle \nonumber \\
&=& \int \mathscr{D}\phi\mathscr{D}\wit{\phi}\; \phi(t,\vec{r}+\vec{r}_0)\phi(s,\vec{r}_0)\, e^{-{\cal J}_0[\phi,\wit{\phi}]-{\cal J}_b[\wit{\phi}]} \nonumber \\
&=& \bigl\langle \phi(t,\vec{r}+\vec{r}_0)\phi(s,\vec{r}_0) e^{-{\cal J}_b[\wit{\phi}]} \bigr\rangle_0 \nonumber \\
&=& \demi \int_{\mathbb{R}^{2d}} \!\!\!\D\vec{R}\D\vec{R}'\: C_0\bigl(\vec{R}-\vec{R}'\bigr)
\left\langle \phi(t,\vec{r}+\vec{r}_0) \phi(s,\vec{r}_0) \wit{\phi}(\vep,\vec{R}) \wit{\phi}(\vep',\vec{R}') \right\rangle_0
\label{corrFT}
\EEA
where we use the decomposition (\ref{actionJdDa}), applied the definition of the deterministic average and then expanded the exponential $e^{-{\cal J}_b[\wit{\phi}]}$ to all orders. 
Because of the Bargman rule (\ref{Bargman}),\footnote{{The Barg\-man su\-per\-se\-lec\-tion rules are valid for deterministic theories 
with a dynamical symmetry group which contains the Galilei group as sub-group \cite{Barg54}. 
Only in this case the relation between averages in the full stochastic theory can be 
reduced in a simple way, like (\ref{corrFT}), to deterministic averages}.} a single term remains which follows from (\ref{actionJdDc}). 
The final formula (\ref{corrFT})} contains the initial correlator, where $\vep,\vep'$ are `initial' time-scales which we shall fix 
later.

\subsection{Schr\"odinger-invariance}

Calculations using (\ref{actionJdD},\ref{detave}) share many aspects of equilibrium calculations.
In what follows we anticipate from subsection~\ref{subsec:2.4} that non-linear contributions to the equation of motion of $\phi$ are irrelevant at late times \cite{Henk25c}.
Then the deterministic action ${\cal J}_0$ can be considered to have the Schr\"odinger group as a dynamical symmetry, as shown explicitly for free fields \cite{Henk03a}
or {the $(1+1)D$ Calogero model} \cite{Shim21}.
The Schr\"odinger Lie algebra is spanned by $\mathfrak{sch}(1)=\langle X_{\pm 1,0}, Y_{\pm\frac{1}{2}}, M_0\rangle$
(for simplicity in a $1D$ notation, see \cite{Henk02,Unte12} for generalisations) where
\BEA
X_n &=& - t^{n+1}\partial_t - \frac{n+1}{2} t^n r\partial_r - (n+1) \delta t^n -\frac{n(n+1)}{4} {\cal M} t^{n-1} r^2 \nonumber \\
Y_m &=& - t^{m+\frac{1}{2}} \partial_r - \left( m + \frac{1}{2}\right) {\cal M} t^{m-\frac{1}{2}} \label{gl:2.6} \\
M_n &=& - t^n {\cal M} \nonumber
\EEA
This is the largest, finite-dimensional Lie algebra of time-space transformations which sends each solution of $\bigl(2{\cal M}\partial_t - \partial_r^2\bigr)\phi=0$ into
another solution, as already known to Jacobi and to Lie. For a historical review see \cite{Duva24} and references therein.
In addition, $\delta$ is the scaling dimension and ${\cal M}>0$ the (non-relativistic) mass of the equilibrium field $\phi$.
The form of the scaling generator $X_0$ implies the dynamical exponent $\mathpzc{z}=2$.
We assume that the order-parameter $\phi$ is {\em quasi-primary} with respect to the Schr\"odinger algebra.\footnote{We follow the standard conformal field-theory terminology of \cite{BPZ}.}
Then the requirement of Schr\"odinger-covariance constrains the form of $n$-point deterministic averages or response functions \cite{Henk94,Henk03a}.
We shall need  here the two-point response function ($\mathscr{R}_0$ is a normalisation constant)
\BEA
\lefteqn{\hspace{-2.5truecm}R(t_a,t_b;r) = \left\langle \phi_a(t_a,\vec{r}) \wit{\phi}_b(t_b,\vec{0})\right\rangle_0
=  \mathscr{R}_0\, \delta({\cal M}_a + \wit{\cal M}_b)\,\delta_{\delta_a,\wit{\delta}_b}\: \Theta(t_a-t_b)} \nonumber \\
&\times&  \bigl( t_a - t_b\bigr)^{-2\delta_a} \exp\left[ - \frac{{\cal M}_a}{2} \frac{\vec{r}^2}{t_a-t_b} \right]
\label{2points}
\EEA
such that response operators $\wit{\phi}$ have negative masses $\wit{\cal M}_b = \wit{\cal M}=-{\cal M}=-{\cal M}_a<0$ and we also notice the constraint $\wit{\delta}=\delta$.
In appendix~A, we shall determine the needed four-point response functions \cite{Golk14,Shim21,Volo09}. 
\BLAU{In the applications we have in mind, we shall consider the scaling limit
\begin{subequations} \label{gl:skal}
\BEQ
s\to\infty \;\; , \;\; \tau = t-s = (y-1)s \to \infty \;\; , \;\; \vec{r}, \vec{R}, \vec{R}' \to \infty
\EEQ
such that the following quantities are kept finite
\BEQ
y = \frac{t}{s} > 1\;\; , \;\; \frac{\vec{r}}{s^{1/2}} \;\; , \;\; \frac{\vec{R}}{s^{1/2}} \;\; , \;\; \frac{\vec{R}'}{s^{1/2}}
\EEQ
\end{subequations}
In this scaling limit (\ref{gl:skal}), we have}
\BEA
\left\langle \phi(t,\vec{r})\phi(s,\vec{0}) \wit{\phi}(0,\vec{R})\wit{\phi}(0,\vec{R}')\right\rangle_0
&\simeq& \bigl(ts\bigr)^{-2\delta} \exp\left[-\frac{\cal M}{2}\frac{\vec{r}^2}{t-s}-{\cal M}\frac{\fR^2+\bR^2}{s}\right] \nonumber \\
&& \times~~ \mathscr{F}^{(2)}\left(\frac{\fR}{s^{1/2}}+\frac{1}{y}\frac{\vec{r}}{s^{1/2}},\frac{\bR}{s^{1/2}}\right)
\label{eq:2.8}
\EEA
with the new space variables $\fR = \demi\bigl( \vec{R} + \vec{R}'\bigr)$ and $\bR = \demi\bigl( \vec{R} - \vec{R}'\bigr)$ and we extended the result from appendix~A to $d$ dimensions.
In addition, $\mathscr{F}^{(2)}$ is an undetermined (differentiable) function of two arguments and $y=t/s$.
This is a special case of the generic form \cite{Golk14,Volo09}.\footnote{Via usual dualisation techniques, this could be generalised
in order to derive explicitly the causality conditions (via Heaviside functions $\Theta$) 
for these response functions \cite{Henk03a,Henk15b}, but is not spelled out in (\ref{eq:2.8},\ref{eq:2.9}).}
For realisation in Fermi gases, see \cite{Beka12,Fuer09,Pal18}.
However, when $t=s>0$, we have the so-called `pairwise equal-time case' \cite{Shim21} and rather obtain, \BLAU{again for the scaling limit (\ref{gl:skal})}
\BEA
\left\langle \phi(s,\vec{r})\phi(s,\vec{0}) \wit{\phi}(0,\vec{R})\wit{\phi}(0,\vec{R}')\right\rangle_0
\simeq s^{-4\delta}\exp\left[-{\cal M}\frac{\fR^2+\bR^2}{s}+{\cal M}\frac{\vec{r}\cdot\fR}{s}\right]
\mathscr{F}^{(1)}\left(\frac{\vec{r}\cdot\bR}{s}\right)
\label{eq:2.9}
\EEA
where $\mathscr{F}^{(1)}$ is an undetermined (differentiable) function of a single argument.
In both (\ref{eq:2.8},\ref{eq:2.9}), the time-scale `$0$' of the response operators $\wit{\phi}$ is meant as a short-hand
for an `initial' time-scale $\vep \ll s,t$ which will be specified later.
In those applications where we shall study finite-size effects \cite{Fish71,Barb83,Suzuki77} in fully finite systems of linear size $N$,
eq.~(\ref{eq:2.8}) will be generalised to
contain a scaling function $\mathscr{F}^{(2,N)}\left(\frac{\fR}{s^{1/2}}+\frac{1}{y}\frac{\vec{r}}{s^{1/2}},\frac{\bR}{s^{1/2}},\frac{t^{1/2}}{N}\right)$ for $t\gg s$, see appendix~B.
Therefore, (\ref{eq:2.8}) can be re-used if one merely replaces the scaling function $\mathscr{F}^{(2)}$ by the function $\mathscr{F}^{(2,N)}$ in (\ref{gl:B5}).

Eq.~(\ref{eq:2.9}) was also derived in the context of the $(1+1)D$ Calogero model \cite{Shim21}, 
with a specific model-dependent expression of the function $\mathscr{F}^{(1)}$.

\subsection{Observables far from equilibrium} \label{subsec:2.3}

Physical ageing occurs far from equilibrium and is not time-translation-invariant. 
Rather than dropping the time-translation generator $X_{-1}$ from the Schr\"odinger algebra, 
we use an inspiration from dynamical symmetries of non-equilibrium systems \cite{Stoi22,Henk23b}
and propose to achieve far-from-equilibrium physics, without standard time-translation-invariance, by the following

\noindent
{\bf Postulate:} \cite{Henk25c} {\it The Lie algebra generator $X_n^{\rm equi}$ of a time-space symmetry of an equilibrium system becomes a symmetry
out-of-equilibrium by the change of representation}
\BEQ \label{gl:hyp}
X_n^{\rm equi} \mapsto X_n = e^{\xi \ln t} X_n^{\rm equi} e^{-\xi \ln t}
\EEQ
{\it where $\xi$ is a dimensionless parameter whose value contributes to characterise the scaling operator $\phi$ on which $X_n$ acts.}
The explicit generators (\ref{gl:2.6}) should be considered as `equilibrium'.

{This is suggestive in the case of critical dynamics at $T=T_c$, where known equilibrium forms of dynamical symmetries \cite{Card85} can be generalised towards a non-equilibrium
representation. The practical success of the procedure \cite{Henk25c,Henk25d} encourages us to re-use the same kind of non-equilibrium representation also in phase-ordering kinetics
for $T<T_c$, although there does not exist an `equilibrium form' of these dynamical symmetries in the ordered phase at equilibrium, 
but where dynamical scaling far-from-equilibrium is well-established \cite{Bray90,Bray94a}.} 

When applied to the dilatation generator $X_0^{\rm equi}$ this merely leads to a modified scaling dimension $\delta_{\rm eff} = \delta-\xi$, since
\begin{subequations} \label{gl:Xgen}
\BEQ \label{gl:X0gen}
X_0^{\rm equi} \mapsto X_0 = -t\partial_t - \frac{1}{\mathpzc{z}}r\partial_r - \bigl(\delta - \xi\bigr)
\EEQ
However, the time-translation generator $X_{-1}^{\rm equi}=-\partial_t$ now becomes
\BEQ \label{gl:X-1gen}
X_{-1}^{\rm equi} \mapsto X_{-1} = -\partial_t + \frac{\xi}{t}
\EEQ
\end{subequations}
Significantly, in this new representation the scaling operators become 
$\Phi(t) = t^{\xi} \phi(t) = e^{\xi \ln t}\phi(t)$ and the equilibrium response functions
(\ref{2points}) are adapted into non-equilibrium ones via
(spatial arguments are suppressed for clarity)\footnote{This depends on the applicability of (\ref{gl:hyp}). 
Different changes of representation are conceivable \cite{Henk25c,Henk25f}.}
\BEA
\bigl\langle \phi_a(t_a)\wit{\phi}_b(t_b)\bigr\rangle_0
&\mapsto & t_a^{\xi_a} t_b^{\wit{\xi}_b} \: \bigl\langle \phi_a(t_a)\wit{\phi}_b(t_b)\bigr\rangle_0
\nonumber \\
\bigl\langle \phi_a(t_a)\phi_b(t_b)\wit{\phi}_c(t_c)\wit{\phi}_d(t_d)\bigr\rangle_0
&\mapsto &
t_a^{\xi_a} t_b^{\xi_b} t_c^{\wit{\xi}_c} t_d^{\wit{\xi}_d} \: \bigl\langle \phi_a(t_a) \phi_b(t_b)\wit{\phi}_c(t_c)\wit{\phi}_d(t_d)\bigr\rangle_0
\label{reponseHE}
\EEA
This means that, beside the mass, we now characterise a {\em non-equilibrium} scaling operator $\phi$ 
by a pair of scaling dimensions $(\delta,\xi)$ and a non-equilibrium response operator
$\wit{\phi}$ by a pair $(\wit{\delta},\wit{\xi})$.
As a consequence of Schr\"odinger-invariance (\ref{2points}), and the Bargman rule (\ref{Bargman}) with $n=m=1$, we have
\BEQ
\delta = \wit{\delta}
\EEQ
but $\xi$ and $\wit{\xi}$ remain independent.

Having {unknowingly} been anticipated long ago \cite{Henk06,Mini12}, 
the postulate (\ref{gl:hyp},\ref{reponseHE}) was studied for two-time auto-responses and auto-correlators in \cite{Henk25,Henk25c} 
{for quenches to $T\leq T_c$} where we explored the requirements of covariance
only under the two generators (\ref{gl:Xgen}). For example, in the two-time auto-correlator 
$C(t,s)=\langle \phi_1\phi_2\rangle=\langle \phi(t)\phi(s)\rangle$ the scaling operator
identity $\phi_1=\phi_2=\phi$ implies for the scaling dimensions $\delta_1=\delta_2=\delta$ and $\xi_1=\xi_2=\xi$. For $t\gg s$, this leads to
\BEQ
C(t,s) =  s^{-2(\delta-\xi)} \bigl( t/s\bigr)^{-(2\delta-\xi)} \mathscr{F}_C(0)
\EEQ
where $\mathscr{F}_C(0)$ is a normalisation constant. Agreement with the expected scaling forms 
(\ref{gl:2},\ref{gl:3a}) implies that, specifically for phase-ordering kinetics
\BEQ \label{gl:Cphaseord}
\frac{\lambda_C}{2} = 2\delta - \xi \;\; , \;\; \delta = \xi
\EEQ
Both of these identities will be needed frequently below.
Here, we shall study if (\ref{reponseHE}) also applies to four-point functions, 
\BLAU{notably for phase-ordering kinetics at $T<T_c$ when the `equilibrium' model is not even scale-invariant.}

\subsection{Irrelevance of non-linearities at long times} \label{subsec:2.4}

The postulated change of representation (\ref{gl:hyp}) leads to another important consequence which we shall need in our calculations.
We are interested in the long-time behaviour of the order-parameter $\phi(t,\vec{r})$ in
phase-ordering kinetics. For long times, 
{it is no longer described by the original equation of motion derived from the action (\ref{actionJdD}), which \BLAU{rapidly becomes} unstable \cite{Bray94a}, but rather} 
should be described by an effective equation of motion, which can be assumed to take the form
\BEQ \label{gl:phi-eff}
\left( \partial_t - \frac{1}{2{\cal M}} \Delta_{\vec{r}} \right) \phi(t,\vec{r}) = g \phi^3(t,\vec{r})
\EEQ
As argued before \cite{Henk25c}, this form is suggestive for the following reasons:
\begin{enumerate}
\item a term linear in $\phi(t,\vec{r})$ on the right-hand-side of (\ref{gl:phi-eff}) would break dynamical scaling
\item a term quadratic in $\phi(t,\vec{r})$ would break the global spin-reversal-invariance {$\phi\mapsto -\phi$}
\item a term cubic in $\phi(t,\vec{r})$ is the lowest-order term which may appear (higher-order terms will lead to corrections to scaling)
\item thermal noise will merely lead to corrections to scaling, hence it can be left out {(noise-dependent correlators are to be found from (\ref{corrFT}))} 
\item the dynamical exponent $\mathpzc{z}=2$ \cite{Bray94b} of systems with short-ranged interactions is included
\end{enumerate}

On the left-hand-side of (\ref{gl:phi-eff}) we have the Schr\"odinger operator 
$\mathscr{S}^{\rm equi}=\partial_t - \frac{1}{2{\cal M}}\Delta_{\vec{r}}$, which in the
new representation according to (\ref{gl:hyp}) will become
\BEQ \label{gl:phi-eff2}
{\mathscr{S}} = e^{\xi \ln t} \mathscr{S}^{\rm equi} e^{-\xi \ln t} = \partial_t - \frac{\xi}{t} - \frac{1}{2{\cal M}} \Delta_{\vec{r}}
\EEQ
and contains an additional $1/t$-potential.\footnote{Such $1/t$-potentials have arisen several times in the literature \cite{Oono88,Maze90,Maze04,Maze06}.}
The original equation $\mathscr{S}^{\rm equi} \phi =g  \phi^3$ then becomes, with the notation of subsection~\ref{subsec:2.3}
\BEQ \label{gl:phi-eff3}
\left( t^{\xi}\, \mathscr{S}^{\rm equi}\, t^{-\xi} \right) \left( t^{\xi}\, \phi \right) = g\, t^{\xi} \left( t^{-\xi}\, {\Phi}\, \right)^{3}
~~\Longrightarrow~~
{\mathscr{S}}\,{\Phi} = \left( \partial_t - \frac{\xi}{t} - \frac{1}{2{\cal M}} \Delta_{\vec{r}} \right) {\Phi} = g\, t^{-2\xi}\, {\Phi}^{\,3}
\EEQ
For phase-ordering kinetics, (\ref{gl:Cphaseord}) implies that $\delta_{\rm eff}=\delta-\xi=0$ such that ${\Phi}$ is dimensionless.

The long-time behaviour of (\ref{gl:phi-eff3}) is governed by the explicit $t$-dependence.
The $1/t$-potential will for large times dominate over against the non-linear term, when 
\BEQ \label{gl:lambda-crit}
2\xi > 1 ~~\Longleftrightarrow~~ \lambda > 1
\EEQ
where we re-used (\ref{gl:Cphaseord}) (and higher-order non-linear terms in (\ref{gl:phi-eff}) are even less relevant).
Hence we have a simple {\bf criterion} of the irrelevance of the non-linearity in the effective equation of motion.
Because of the well-known {\sc yrd}-bound $\lambda\geq d/2$ \cite{Yeun96a}, the criterion (\ref{gl:lambda-crit}) holds true for all $d>2$. For $d=2$, one has
typically $\lambda\approx 1.25 >1$ (see \cite{Henk10} and refs. therein) and the criterion (\ref{gl:lambda-crit}) is satisfied as well.\footnote{A counter-example
occurs in the $1D$ {\sc tdgl}-equation quenched to $T=0$, where $\lambda_C=0.600616\ldots$ \cite{Bray95}.}
{It follows that, although the effective equation (\ref{gl:phi-eff}) of phase-ordering need not be Schr\"odinger-invariant,
we can nevertheless apply the Schr\"odinger symmetry to deduce its long-time behaviour from the linear part of the equation 
(\ref{gl:phi-eff3}), with the additional $1/t$-potential.} 
\BLAU{But this is not meant to suggest that exponents such as $\lambda, a, \ldots$ could be found from a linear equation. 
Rather, any microscopic calculation of their values must use the full non-linear equations of motion, e.g. \cite{Maze90,Maze04,Maze06,Cala05,Taeu14}. We shall treat them,
for our purposes, as free parameters and substitute their numerical values as taken from the literature.}

In this discussion,\footnote{This argument does not apply to equations $\mathscr{S}^{\rm equi}\phi= g \phi \bigl( \phi\wit{\phi}\,\bigr)^m$,
since in phase-ordering $\xi+\wit{\xi}=0$.}  the coupling $g$ was assumed to be dimensionless.
It can be shown that for a dimensionful coupling $g$ in (\ref{gl:phi-eff}),
the criterion (\ref{gl:lambda-crit}) is still applicable \cite{Stoi05,Henk25c}. {Use of a dimensionful
coupling $g$ permits to circumvent the constraint $\delta=\frac{d}{4}$.}

\section{Calculation of correlators} \label{sec:3}

In physical ageing, observables include both response and correlation functions. The general line of argument is as follows.
First, response functions will be assumed to transform covariantly under local scale-transformations.
Second, due to the relation (\ref{reponseHE}) which related the responses {\em at} equilibrium to their non-equilibrium analogues, it can be
shown that in the effective equation of motion (\ref{gl:phi-eff3}) the non-linear terms are 
irrelevant for the leading long-time behaviour, if the auto-correlation exponent obeys the criterion $\lambda>1$.
Third, for phase ordering-kinetics where the dynamical exponent $\mathpzc{z}=2$ \cite{Bray94b}, the relevant dynamical symmetry is given  by Schr\"odinger-invariance.
A quasi-primary scaling operator $\phi$ is characterised by the pair $(\delta,\xi)$, besides its mass ${\cal M}>0$, and similarly for its response operator $\wit{\phi}$, with
negative mass $\wit{\cal M}<0$. 
Combining (\ref{reponseHE},\ref{2points}) for the two-time response we have for $t>s$
\begin{subequations} \label{2reponse}
\begin{align} \label{2reponseR}
R(t,s;r) = \left\langle {\phi}(t,\vec{r}) {{\wit{\phi}}}(s,\vec{0})\right\rangle
= \mathscr{R}_0 \, s^{-1-a} \left( \frac{t}{s}\right)^{1+a'-\lambda_R/2}
\left(\frac{t}{s} -1 \right)^{-1-a'} \exp\left[ -\frac{\cal M}{2} \frac{r^2}{t-s} \right]
\end{align}
where the exponents $a,a',\lambda_R$ of ageing are related to $\delta,\xi,\wit{\xi}$ as follows
\BEQ \label{2reponse-exp}
\frac{\lambda_R}{2} = 2\delta -\xi \;\; , \;\; 1+a = 2\delta -\xi - \wit{\xi} \;\; , \;\; a'-a = \xi +\wit{\xi}
\EEQ
and ${\cal M},\mathscr{R}_0$ are non-universal, dimensionful 
constants.\footnote{We stress that the gaussian spatial form (\ref{2reponseR}) has been confirmed numerically in the non-gaussian
$2D/3D$ Glauber-Ising models quenched to $0<T<T_c$ \cite{Henk03}.}
In phase-ordering, since in all known models one finds $a'-a=0$ (see \cite{Henk10,Henk25c} and refs. therein)
we have the additional information
\BEQ \label{2response-xi}
\xi = - \wit{\xi}  \;\; ;\;\; \mbox{\rm ~~if~~ $0<T < T_c$}
\EEQ
\end{subequations}
In appendix~C, we study further extensions of representations of the Schr\"odinger algebra by additional dimensionful variables (such as an initial magnetisation $m_0$ or a
finite-size $1/N$) and show that these extensions will not lead to further restrictions to the form of the two-time response function $R(t,s;\vec{r})$.
It follows that all previous results on the two-time responses can be taken over without modification and that the model evidence presented in \cite{Henk25c} is still valid.
Eqs.~(\ref{2reponse}) contain all required information on $R(t,s;\vec{r})$ and is in full agreement with the scaling expectation (\ref{gl:local}).

This is different for the correlation functions. The formal treatment, assuming covariance under (\ref{gl:Xgen}),
only holds for the two-time correlators but cannot be extended to the single-time ones \cite[Appendix B]{Henk25c} 
such that the method is probably best considered as being merely heuristic.
To re-derive these results, we shall start here from the assumption of the covariance of the four-point
response $\langle\phi\phi\wit{\phi}\wit{\phi}\rangle_0$. This permits to include a treatment of the single-time correlators as well.
In this way, we achieve a consistent physical treatment of all observables.

\subsection{Two-time auto-correlator} \label{subsec:2t}

It is enough to collect the preparations made in the previous section~\ref{sec:2}. 
Combining (\ref{corrFT},\ref{eq:2.8},\ref{reponseHE}) and setting $\vec{r}=\vec{0}$, we have
\BEA
C(ys,s;\vec{0}) &=& \!\int_{\mathbb{R}^{2d}} \!\!\D\fR\D\bR\: C_0\bigl(2\bR\bigr)\, 
y^{-2\delta+\xi} s^{2\xi-4\delta} \vep^{2\wit{\xi}}\:e^{-\frac{{\cal M}}{s}\bigl(\fR^2+\bR^2\bigr)}\,
\mathscr{F}^{(2)}\left(\fR\frac{(s-1)^{1/2}}{s},\bR\frac{(s-1)^{1/2}}{s}\right)
\nonumber \\
&\simeq& y^{-2\delta+\xi}\, s^{2(\xi-\delta)}\,s^{-2\delta+d} \vep^{2\wit{\xi}}
\int_{\mathbb{R}^{2d}} \!\D \vec{U}\D\lap{\vec{U}}\: C_0\bigl(2 \lap{\vec{U}} s^{1/2}\bigr) 
e^{-{\cal M}\bigl(\vec{U}^2+\lap{\vec{U}}^2\bigr)}\, \mathscr{F}^{(2)}\bigl(\vec{U}, \lap{\vec{U}}\bigr)
\label{gl:3.2}
\EEA
where in the second line, we first let $s\gg 1$ and then changed the integration variables.
In what follows, we shall always assume that the initial correlator $C_0\bigl(\vec{R}\bigr)$ as well as
the scaling function $\mathscr{F}^{(2)}\bigl(\vec{U},\lap{\vec{U}}\bigr)$ are such that in the limit of large waiting times $s\gg 1$
\BEQ
\int_{\mathbb{R}^{2d}} \!\D \vec{U}\D\lap{\vec{U}}\:
C_0\bigl(2 \lap{\vec{U}} s^{1/2}\bigr)\: e^{-{\cal M}\bigl(\vec{U}^2+\lap{\vec{U}}^2\bigr)}\, \mathscr{F}^{(2)}\bigl(\vec{U}, \lap{\vec{U}}\bigr)
\longrightarrow \mathscr{C}^{(2)}_{\infty}
\label{gl:3.3}
\EEQ
tends towards a finite, non-vanishing constant $\mathscr{C}^{(2)}_{\infty}$.
Furthermore, as inspired by the studies in \cite{Zipp00,Andr06}, we admit that the `initial' time-scale at the beginning
of the scaling regime is related to the waiting time $s$ as
\BEQ \label{gl:3.4}
\vep \simeq \vep_0\,  s^{\zeta_p}
\EEQ
where $\zeta_p$ is a new exponent supposed to describe the beginning of the scaling regime. With these assumptions,
the leading large-time behaviour (\ref{gl:3.2}) of the two-time auto-correlator becomes
\BEQ \label{gl:3.5}
C(ys,s) = y^{-(2\delta-\xi)}\, s^{2(\xi-\delta)}\,s^{d-2\delta+2\zeta_p\wit{\xi}}\: \vep_0\, \mathscr{C}^{(2)}_{\infty}
\EEQ
This already reproduces the algebraic behaviour (\ref{gl:3a}) of the two-time correlator for $y=t/s\gg 1$, as expected from figure~\ref{fig1}d.
Comparison with (\ref{2reponse-exp}) shows that $\frac{\lambda_C}{2}=2\delta-\xi=\frac{\lambda_R}{2}$ 
and thus proves the anticipated exponent equality (\ref{lambda}).
In phase-ordering kinetics, we have the identities (\ref{gl:Cphaseord}) and (\ref{2response-xi}), that is $\delta=\xi=-\wit{\xi}$.
The scaling (\ref{gl:3.5}) becomes $s$-independent, as expected from (\ref{gl:2}), if we have the scaling relation
\BEQ \label{gl:3.6}
2\delta = \frac{d}{1+\zeta_p} = \lambda
\EEQ
As we shall see below, condition (\ref{gl:3.6}) will arise frequently 
in different physical situations.\footnote{The waiting time $s$ 
must be large enough that the system is in the scaling regime.} 
\BLAU{The required existence of the scaling relation (\ref{gl:3.6}) is an important difference with respect to what occurs for non-equilibrium critical dynamics at $T=T_c$.}
It underscores the non-trivial nature of the auto-correlation exponent $\lambda$.

For the passage exponent, one has obviously $\zeta_p\geq 0$ and also $\zeta_p\leq 1$ since the ageing regime cannot start later than at the waiting time $s$ itself. 
Together with (\ref{gl:3.6}) this gives the bounds
\BEQ
\frac{d}{2} \leq \lambda \leq d
\EEQ
These merely reconfirm long-standing bounds on $\lambda$  from the literature \cite{Fish88a,Yeun96a} and serve to us as a consistency check of the present treatment.

Since the change of representation (\ref{gl:hyp}) is supposed to be a leading large-time expression, 
this should mean that the representation (\ref{gl:3.5}) merely gives the leading long-time behaviour.

\subsection{Single-time correlator}

Now, we set $t=s$, combine  (\ref{corrFT},\ref{eq:2.9},\ref{reponseHE}) and have
\BEA
\lefteqn{ C(s;r) = \int_{\mathbb{R}^{2d}}\!\D\fR\D\bR\: C_0\bigl(2\bR\bigr)\: 
e^{-\frac{{\cal M}}{s}\bigl(\fR^2+\bR^2\bigr)+\frac{\cal M}{s}r\,\fR}\, s^{2\xi-4\delta}\vep^{2\wit{\xi}}
\mathscr{F}^{(1)}\left( \frac{\vec{r}\cdot\bR}{s} \right) }
\nonumber \\
&=& s^{2\xi-4\delta+2\wit{\xi}\zeta_p} \int_{\mathbb{R}^d} \!\D\fR\: 
e^{-\frac{\cal M}{s}\bigl( \fR^2 -2\fR\frac{r}{2} +\bigl(\frac{r}{2}\bigr)^2\bigr)} e^{-\frac{\cal M}{s}\frac{r^2}{4}}\,
\int_{\mathbb{R}^d}\!\D\bR\: C_0\bigl(2\bR\bigr)\, 
e^{-\frac{\cal M}{s} \bR^2}\, \mathscr{F}^{(1)}\left( \frac{\vec{r}\cdot\bR}{s} \right)
\nonumber \\
&=& e^{-\frac{\cal M}{4}\frac{r^2}{s}}\, s^{2\xi-4\delta+\frac{d}{2}-2\xi\zeta_p}
\int_{\mathbb{R}^d}\!\D\bR\: C_0\bigl(2\bR\bigr)\: 
e^{-\frac{\cal M}{s} \bR^2}\, \mathscr{F}^{(1)}\left( \frac{\vec{r}}{s^{1/2}}\cdot\frac{\bR}{s^{1/2}} \right)
\nonumber \\
&=& e^{-\frac{\cal M}{4}\frac{r^2}{s}}\, s^{-2\delta+d-2\zeta_p\delta}
\int_{\mathbb{R}^d}\!\D\lap{\vec{U}}\: C_0\bigl(2\lap{\vec{U}} s^{1/2}\bigr)\: 
e^{-{\cal M}\lap{\vec{U}}^2}\, \mathscr{F}^{(1)}\left( \frac{\vec{r}}{s^{1/2}}\cdot \lap{\vec{U}} \right)
\EEA
In carrying this out, we introduced the `initial' time estimate (\ref{gl:3.4}), 
then completed a square in the $\fR$-integration in the second line to be able to carry out this integration,
and finally changed the last integration variable. Using $\delta=\xi=-\wit{\xi}$ 
and applying the scaling relation (\ref{gl:3.6}) once more, we finally have
\BEQ \label{gl:3.9}
C(s;r) = e^{-\frac{\cal M}{4}\frac{r^2}{s}}\:
\int_{\mathbb{R}^d}\!\D\lap{\vec{U}}\: C_0\bigl(2\lap{\vec{U}}\, s^{1/2}\bigr)\: 
e^{-{\cal M}\lap{\vec{U}}^2}\, \mathscr{F}^{(1)}\left( \frac{\vec{r}}{s^{1/2}}\cdot \lap{\vec{U}} \right)
\EEQ
and the scaling variable $r/\sqrt{s\,}$ naturally emerges.
If the last integral has a finite limit for $s\gg 1$, we have reproduced once more the scaling form (\ref{gl:2}), now for $t=s$.
We can identify the scaling function $F_C(1,r/\sqrt{s\,}\,)$ which indeed comes out spatially rotation- and translation-invariant, as expected.

An explicit computation of the scaling function $F_C(1,r/\sqrt{s\,}\,)$ must await stronger information on the function $\mathscr{F}^{(1)}$ than is currently available.
But if a limited analytic expansion of $\mathscr{F}^{(1)}$ for small arguments is possible, we would find an expansion of $C(s;r)$ for small $r$
\BEQ
C\bigl(s;|\vec{r}|\bigr) \simeq \exp\left[ {-\frac{\cal M}{4}\frac{r^2}{s}}\right]
\int_{\mathbb{R}^d}\!\D\lap{\vec{U}}\: C_0\bigl(2\lap{\vec{U}} s^{1/2}\bigr)\: e^{-{\cal M}\lap{\vec{U}}^2}\,
\left( \mathscr{F}_0^{(1)} + \mathscr{F}_1^{(1)}\, \lap{\vec{U}}\cdot \frac{\vec{r}}{s^{1/2}} + \ldots \right)
\EEQ
If the corresponding integrals have finite limits for $s\gg 1$ and if $\mathscr{F}_1^{(1)}<0$, 
this would reproduce the typical small-distance behaviour of the
single-time correlator for a scalar order-parameter, with a cusp at $r=0$,
of the form $C(s;|\vec{r}|)\simeq C_0 - C_1 \frac{|\vec{r}|}{s^{1/2}} +\ldots$, illustrated in figure~\ref{fig1}b and
as predicted by {\sc Porod}'s law \cite{Poro51,Bray94a,Gall12}, see also appendix~D.\footnote{\BLAU{A nice recent experimental illustration of the ageing phenomenon in phase-ordering kinetics, 
especially on {\sc Porod}'s law and more generally on the shape of single-time and two-time correlators, 
arises in liquid crystals \cite{Alme21}.}}
On the other hand, if $\mathscr{F}_1^{(1)}=0$ vanishes, then the gaussian prefactor will give a rounded-off contribution for $r\ll 1$. 
There are many illustrations for this in the literature, e.g. \cite{Chris19}, and for a classic example see \cite[fig. 14]{Bray94a}.

Remarkably, single-time and two-time correlators are treated on the same conceptual basis,
namely the covariance of the four-point response function $\langle\phi\phi\wit{\phi}\wit{\phi}\rangle$.

\subsection{Structure factor}

Starting from the definition of the structure factor, we have using (\ref{gl:3.9})
\BEA
\wht{S}(s;\vec{q}) &:=& \int_{\mathbb{R}^d} \!\D\vec{r}\: e^{-\II \vec{q}\cdot\vec{r}}\: C(s;\vec{r}) \nonumber \\
&=& \int_{\mathbb{R}^d} \!\D\vec{r}\: e^{-\II\vec{q}\cdot\vec{r}}\: e^{-\frac{\cal M}{4}\frac{\vec{r}^2}{s}}
\int_{\mathbb{R}^d} \!\D\lap{\vec{U}}\: C_0\bigl(2\lap{\vec{U}}\, s^{1/2}\bigr)\: 
e^{-{\cal M}\lap{\vec{U}}^2}\, \mathscr{F}^{(1)}\left( \frac{\vec{r}}{s^{1/2}}\cdot \lap{\vec{U}} \right)
\nonumber \\
&=& s^{d/2} \int_{\mathbb{R}^{2d}} \!\D\vec{u}\D\lap{\vec{U}}\: e^{-\II \bigl(\vec{q}\,s^{1/2}\bigr)\cdot\vec{u}}\:
C_0\bigl(2\lap{\vec{U}}\, s^{1/2}\bigr)\: 
e^{-\frac{\cal M}{2} \vec{u}^2 -{\cal M}\lap{\vec{U}}^2}\: \mathscr{F}^{(1)}\left( \vec{u}\cdot \lap{\vec{U}} \right)
\nonumber \\
&=& s^{d/2} \mathpzc{g}\left( \vec{q} s^{1/2} \right) \:=\: \ell(s)^d \bar{\mathpzc{g}}\bigl( \vec{q} \ell(s) \bigr)
\EEA
which has the required scaling form \cite{Bray94a}, with scaling functions $\mathpzc{g}$ or $\bar{\mathpzc{g}}$, 
and the length scale $\ell=\ell(s)\sim s^{1/2}$,
on the basis of the same assumptions on the functions $C_0\bigl(\vec{R}\bigr)$ and $\mathscr{F}^{(1)}$ as before.

In the limit  $|\vec{q}|\to\infty$, this should be compatible with {\sc Porod}'s law \cite{Poro51,Bray94a,Gall12}, see appendix~D for background.
Since for a large momentum, the main contribution to the integral comes from $|\vec{u}|$ small,
consider the expansion $\mathscr{F}^{(1)}\bigl(\vec{u}\cdot\lap{\vec{U}}\bigr)\simeq \mathscr{F}_0^{(1)} + \mathscr{F}_1^{(1)} \vec{u}\cdot\lap{\vec{U}} +\ldots$.
Then, with $\vec{Q}=\vec{q} s^{1/2}$
\BEA
\mathpzc{g}(\vec{Q}) &\simeq& \int_{\mathbb{R}^{2d}} \!\D\vec{u}\D\lap{\vec{U}}\: e^{-\II \vec{Q}\cdot\vec{u}}\:
C_0\bigl(2\lap{\vec{U}}\, s^{1/2}\bigr)\: e^{-\frac{\cal M}{2} \vec{u}^2 -{\cal M}\lap{\vec{U}}^2}\: 
\left( \mathscr{F}_0^{(1)} +\mathscr{F}_1^{(1)} \vec{u}\cdot\lap{\vec{U}} +\ldots\right)
\nonumber \\
&=& \mathscr{F}_0^{(1)} \int_{\mathbb{R}^d} \!\D\vec{u}\: e^{-\II \vec{Q}\cdot\vec{u} -\frac{\cal M}{2} \vec{u}^2}
\int_{\mathbb{R}^d} \!\D\lap{\vec{U}}\: C_0\bigl(2\lap{\vec{U}}\, s^{1/2}\bigr)\: e^{-{\cal M}\lap{\vec{U}}^2} \nonumber \\
&& ~~+ \mathscr{F}_1^{(1)} \int_{\mathbb{R}^{2d}} \!\D\vec{u}\D\lap{\vec{U}}\: e^{-\II \vec{Q}\cdot\vec{u}}\:
C_0\bigl(2\lap{\vec{U}}\, s^{1/2}\bigr)\: e^{-\frac{\cal M}{2} \vec{u}^2 -{\cal M}\lap{\vec{U}}^2}\: \vec{u}\cdot\lap{\vec{U}}  +\ldots
\nonumber \\
&=& \mathscr{F}_0^{(1)} \underbrace{~~e^{-\frac{\vec{Q}^2}{2{\cal M}}}~~}_{\longrightarrow~ 0}
\underbrace{~\int_{\mathbb{R}^d} \!\D\vec{u}\: 
e^{-\frac{\cal M}{2}\bigl[ \vec{u}^2 + 2\II \frac{\vec{Q}\cdot\vec{u}}{\cal M}-\frac{\vec{Q}^2}{{\cal M}^2}\bigr]}
\int_{\mathbb{R}^d} \!\D\lap{\vec{U}}\: C_0\bigl(2\lap{\vec{U}}\, s^{1/2}\bigr)\: 
e^{-{\cal M}\lap{\vec{U}}^2}~}_{=~ \mbox{\rm\fns cste.}} \nonumber \\
&& ~~+Q^{-d-1} \mathscr{F}_1^{(1)} \int_{\mathbb{R}^{2d}} \!\D\vec{u}\D\lap{\vec{U}}\:
C_0\bigl(2\lap{\vec{U}}\, s^{1/2}\bigr)\: 
e^{-\frac{\cal M}{2} \frac{\vec{u}^2}{\vec{Q}^2} -{\cal M}\lap{\vec{U}}^2}\:\vec{u}\cdot\lap{\vec{U}}\: e^{-\II \vec{Q}_e\cdot\vec{u}}
\nonumber \\
&\stackrel{Q\gg 1}{\simeq}& Q^{-d-1}\: \mbox{\rm cste.}
\EEA
where $\vec{Q}=Q\,\vec{Q}_e$  with the fixed vector $\vec{Q}_e$ of unit length $|\vec{Q}_e|=1$. 
Changing the integration variable $\vec{u}\mapsto \vec{u}-\II\frac{\vec{Q}}{\cal M}$,
we see that the first term in the third line vanishes exponentially fast in the $Q=|\vec{Q}|\to\infty$ limit.
The only remaining term reproduces {\sc Porod}'s law (\ref{gl:D2}) for a scalar order-parameter. 
If the first non-vanishing and non-constant term in the function
$\mathscr{F}^{(1)}$ is of order $n$, one would obtain a generalised {\sc Porod}'s law \cite{Bray92a,Bray94a}, 
see (\ref{gl:D2}) and \cite[fig. 15]{Bray94a} for classic illustrations.

\subsection{Fully finite systems} \label{subsec:fss}

We now consider a fully finite system, say in a hypercubic geometry with a side of linear length $N$. Then the auto-correlator, for values $y=t/s\gg 1$ should cross over to a plateau
\cite{Henk23,Henk25c} whose height $C_{\infty}^{(2)}$ will scale with $s$ and with $N$. We can repeat the same steps as in subsection~\ref{subsec:2t}, 
\BLAU{notably re-use the scaling relation (\ref{gl:3.6})} and merely have to replace the
the scaling function in (\ref{eq:2.8}), of the four-point function, by the corresponding scaling function (\ref{gl:B5}) from appendix~B. For $t=ys\gg s>0$, the
two-time auto-correlator can be written in the form
\BEA
C\left(ys,s;\vec{0};\frac{1}{N}\right) &=& y^{-\delta} \int_{\mathbb{R}^{2d}} \!\D\vec{U}\D\lap{\vec{U}}\: C_0\bigl(2\lap{\vec{U}} s^{1/2}\bigr)\:
e^{-{\cal M}\bigl(\vec{U}^2+\lap{\vec{U}}^2\bigr)} \mathscr{F}^{(2,N)}\left( \vec{U},\lap{\vec{U}},\frac{\bigl( ys\bigr)^{1/2}}{N}\right)
\nonumber \\
&\simeq& y^{-\delta} \mathscr{F}_C\left( \frac{N}{\bigl( ys\bigr)^{1/2}}\right)
\EEA
\BLAU{where the finite-size scaling form in the second line holds in the scaling limit (\ref{gl:skal}). Recall that $\delta=\lambda/2$.} 
It only remains to discuss the \BLAU{expected finite-size scaling behaviour of $\mathscr{F}_C$ or equivalently the} 
dependence on the third scaling variable of the scaling function $\mathscr{F}^{(2,N)}$. 
Clearly, for $N\gg t^{1/2}=\bigl( ys\bigr)^{1/2}$, the system will behave as being spatially infinite. 
\BLAU{In that limit, $\mathscr{F}_C$ should become a constant and}
the scaling function $\mathscr{F}^{(2,N)}\bigl(.,.,\mathfrak{u}\bigr)$ \BLAU{is expected to become} independent of $\mathfrak{u}$. On the other hand, for finite systems one
expects $N\lesssim t^{1/2}$ such that \BLAU{$\mathscr{F}_C(\mathfrak{u})\sim \mathfrak{u}^{-\lambda}$ or equivalently} 
$\mathscr{F}^{(2,N)}\bigl(.,.,\mathfrak{u}\bigr)\sim \mathfrak{u}^{\lambda}$. 
Then the plateau height \BLAU{$C_{\infty}^{(2)}=\lim_{s\to\infty} C\bigl(ys,s;\vec{0};\frac{1}{N}\bigr)$} should scale as
\BEQ
C_{\infty}^{(2)} \sim \left( \frac{t}{s}\right)^{-\lambda/2} \left( \frac{t^{1/2}}{N}\right)^{\lambda} \sim N^{-\lambda} s^{\lambda/2}
\EEQ
and in particular, we should have the finite-size scaling behaviour 
\BEQ \label{gl:3.15}
C_{\infty}^{(2)} \sim \left\{
\begin{array}{ll} N^{-\lambda}  & \mbox{\rm if $s$ is kept fixed} \\
                  s^{\lambda/2} & \mbox{\rm if $N$ is kept fixed}
\end{array} \right.
\EEQ
which reproduce that given in \cite{Henk25c} for the special case of quenches to $0<T<T_c$ (hence $b=0$) and for $\mathpzc{z}=2$.

Available tests of this in specific models have been discussed in detail in \cite{Henk25c}.

We have not identified any spectacular finite-size effect in the single-time correlator.

\subsection{Global correlators}

The global two-time correlator for $t>s$ is simply obtained by integrating the two-time correlator $C(t,s;\vec{r})$ with respect to $\vec{r}$.
Combining (\ref{corrFT},\ref{eq:2.8},\ref{reponseHE}) leads to
\BEA
\lefteqn{ \wht{C}(t,s;\vec{0}) = \int_{\mathbb{R}^d} \!\D\vec{r}\:  C(t,s;\vec{r}) } \nonumber \\
&=& \int_{\mathbb{R}^d} \!\D\vec{r} \int_{\mathbb{R}^{2d}} \!\D\fR\D\bR\: a\big(2\bR\bigr)\: (ts)^{\xi}\, \vep^{2\wit{\xi}}\, (ts)^{-2\delta}
\nonumber \\
&& \times \exp\left[ -\frac{\cal M}{2}\frac{\vec{r}^2}{t-s} -{\cal M}\bigl(\fR^2+\bR^2\bigr)\right]
\mathscr{F}^{(2)}\left( \frac{\fR (t-s) +\vec{r} s}{((t-s)t s)^{1/2}}, \bR \frac{(t-s)^{1/2}}{(t s)^{1/2}} \right)
\nonumber \\
&=& (t-s)^{d/2} (ts)^{-\delta} s^{d-2\zeta_p \delta}
\int_{\mathbb{R}^{3d}} \!\D\vec{u}\D\vec{U}\D\lap{\vec{U}}\: C_0\bigl(2\lap{\vec{U}} s^{1/2}\bigr)\: 
e^{-\frac{\cal M}{2} \vec{u}^2 -{\cal M}\bigl(\vec{U}^2+\lap{U}^2\bigr)}\,
\nonumber \\
&& \times
\mathscr{F}^{(2)}\left( \vec{U}\left(\frac{t-s}{t}\right)^{1/2} +\vec{u}\left(\frac{s}{t}\right)^{1/2},\lap{\vec{U}}\left(\frac{t-s}{t}\right)^{1/2} \right)
\nonumber \\
&\simeq&  s^{d/2} \left( \frac{t}{s}\right)^{\Theta} \underbrace{~\int_{\mathbb{R}^{3d}} \!\D\vec{u}\D\vec{U}\D\lap{\vec{U}}\:
C_0\bigl(2\lap{\vec{U}} s^{1/2}\bigr)\: e^{-\frac{\cal M}{2} \vec{u}^2 -{\cal M}\bigl(\vec{U}^2+\lap{U}^2\bigr)}\,
\mathscr{F}^{(2)}\left(\vec{U}+\vec{u}y^{-1/2},\lap{\vec{U}}\right)~}_{=~\mbox{\rm\fns cste.}} ~~~~~
\label{gl:3.16}
\EEA
where in the last line we let $t\gg s$, used as before that $\delta=\xi=-\wit{\xi}$ 
and also the scaling relation (\ref{gl:3.6}) about $\zeta_p$.
We also assume that the last integral in (\ref{gl:3.15}) converges to a finite non-zero constant in the $s\gg 1$ limit. 
{In particular, the global correlator (\ref{gl:3.16})
with the initial state scales as $\wht{C}(t,0)\sim t^{\Theta}$ where the} {\em slip exponent}
\BEQ \label{gl:3.17}
\Theta = \demi \bigl( d - \lambda \bigr)
\EEQ
is given by {the extension to $0<T<T_c$ of} the {\sc Janssen-Schaub-Schmittmann} ({\sc jss}) 
{critical-point} scaling relation \cite{Jans89}, for $\mathpzc{z}=2$, as expected.\footnote{Of course, the values of $\Theta,\lambda,\mathpzc{z}$ are in general different for $0<T<T_c$ and $T=T_c$.}
For quenches onto the critical point $T=T_c$, the original {\sc jss}-relation has been the conceptual basis of a whole field of studies on non-equilibrium critical
dynamics, called `short-time dynamics', since it is not necessary to carry out simulation to extremely long times, see \cite{Alba11,Zhen98} for classical reviews.
Eq.~(\ref{gl:3.17}) could serve the same purpose in phase-ordering kinetics after a quench into $T<T_c$. 

For equal times $t=s$, we might use the combination of (\ref{corrFT},\ref{eq:2.9},\ref{reponseHE}) and find for the squared magnetisation
\BEA
\left\langle m^2(s)\right\rangle &=& \wht{C}(s,s;\vec{0}) =\int_{\mathbb{R}^d} e^{-\frac{\cal M}{4}\frac{\vec{r}^2}{s}}
\int_{\mathbb{R}^d} \!\D\lap{\vec{U}}\: C_0\bigl(2\lap{\vec{U}} s^{1/2}\bigr)\: 
e^{-{\cal M}\lap{\vec{U}}^2}\: \mathscr{F}^{(1)}\left( \frac{\vec{r}}{s^{1/2}}\cdot \lap{\vec{U}}\right)
\nonumber \\
&=& s^{d/2} \underbrace{~\int_{\mathbb{R}^{2d}} \!\D\vec{u}\D\lap{\vec{U}}\: C_0\bigl(2\lap{\vec{U}} s^{1/2}\bigr)\:
\exp\left[-\frac{\cal M}{4} \vec{u}^2 -{\cal M}\lap{\vec{U}}^2\right] \mathscr{F}^{(1)}\left( \vec{u}\cdot\lap{\vec{U}}\right)~}_{=~\mbox{\rm\fns cste.}}
\label{gl:3.18}
\EEA
and with the usual assumption that the last integral converges to a finite, non-zero constant, 
we recover the scaling $\langle m^2(s)\rangle \sim s^{d/2}$ \cite{Jank23} \BLAU{and well-tested in simulations}.
Of course, one may obtain this scaling also from (\ref{gl:3.16}) by taking the $t\to s$ limit.

To finish, we discuss the finite-size scaling of the global auto-correlator in a fully finite system of linear size $N$.
As above, we expect that the global correlator should converge towards a plateau of height $\wht{C}_{\infty}^{(2)}$
when $\ell(t)\approx N$ but $\ell(s)\ll N$. Generalising (\ref{gl:3.16}) we have, for $t\gg s$
\BEA
\lefteqn{ \wht{C}\left(t,s;\vec{0};\frac{1}{N}\right) = \int_{\mathbb{R}^d} \!\D\vec{r}\: C\left(t,s;\vec{r};\frac{1}{N}\right) }
\nonumber \\
&=& s^{d/2} \left(\frac{t}{s}\right)^{\Theta} \int_{\mathbb{R}^{3d}} \!\D\vec{u}\D\vec{U}\D\lap{\vec{U}}\: C_0\bigl(2\lap{\vec{U}} s^{1/2}\bigr)\:
e^{-\frac{\cal M}{2}\vec{u}^2-{\cal M}\bigl(\vec{U}^2+\lap{U}^2\bigr)}\, 
\mathscr{F}^{(2,N)}\left(\vec{U}+\vec{u}\frac{s}{t},\lap{\vec{U}},\frac{t^{1/2}}{N}\right)
\nonumber \\
&\sim & s^{d/2-\Theta} N^{2\Theta}
\label{gl:3.19}
\EEA
and use of course the scaling relation (\ref{gl:3.17}). The phenomenological discussion of the limits $N\gg t^{1/2}$ and $N\lesssim t^{1/2}$
in the scaling function $\mathscr{F}^{(2,N)}$ and the scaling of the plateau
$\wht{C}_{\infty}^{(2)}$ is as before in subsection~\ref{subsec:fss} and leads to the scaling in the last line of (\ref{gl:3.19}). 
More explicitly the plateau height scales as
\BEQ \label{gl:3.20}
\wht{C}_{\infty}^{(2)} \sim \left\{
\begin{array}{ll} N^{2\Theta}    & \mbox{\rm if $s$ is kept fixed} \\
                  s^{d/2-\Theta} & \mbox{\rm if $N$ is kept fixed}
\end{array} \right.
\EEQ
as predicted before for $0<T<T_c$ and $\mathpzc{z}=2$ \cite{Henk25c}.

\section{Conclusions} \label{sec:4}

Our investigations \BLAU{on the scaling of correlators in phase-ordering kinetics, after a quench from a high-temperature initial state into the ordered phase at $T<T_c$, 
showed that their scaling forms can be derived from a conceptually clean basis which only admits
the Schr\"odinger-covariance of multi-point response functions. 
This allows to avoid previous difficulties when the covariance of {\em both} two-time responses and correlators were assumed
under dynamical dilatations and generalised time-translations \cite{Henk25c},\footnote{For generic exponents
$\mathpzc{z}\ne 2$ this remains the only line of reasoning available. Also, the required calculations are much shorter.} 
but which does not allow the treatment of single-time correlators.}
 
\BLAU{The present approach,} based on (\ref{corrFT}), has shown how the four-point response function 
$\langle\phi\phi\wit{\phi}\wit{\phi}\rangle$ can be used to obtain \BLAU{simultaneously} information on the single-time and two-time correlators. 
The form of the \BLAU{far-from-equilibrium} response functions in turn was derived from a new non-equilibrium representation of the Lie algebra generators
\BEQ \label{gl:4.1}
X^{\rm equi} \mapsto X = e^{W(t)}\: X^{\rm equi}\: e^{-W(t)} \;\; , \;\; W(t) = \xi \ln t
\EEQ
where the specific choice of $W(t)$ investigated here appears to be the one which reproduces the expected non-equilibrium (ageing) phenomenology (\ref{gl:local}), at least in the
most frequently studied models. The long-standing Schr\"odinger Lie algebra $\mathfrak{sch}(1)$ \cite{Duva24} was used as the underlying dynamical symmetry, 
\BLAU{as is suggested by the well-known fact that $\mathpzc{z}=2$ for model-A-type dynamics without a macroscopic conservation law \cite{Bray94b}.}  
\BLAU{Transcribing this way the results of `equilibrium' Schr\"odinger-invariance with generators (\ref{gl:2.6}) to a far-from-equilibrium situation apparently also works in
situations when the `equilibrium' system itself is not dynamically scale-invariant.
All other presently known tests refer to the context of non-equilibrium critical dynamics at $T=T_c$ \cite{Henk25d,Henk25f}, in systems where in addition $\mathpzc{z}=2$}.

\BLAU{Our approach strives to} constrain the form of the response functions from their assumed covariance under a non-equilibrium \BLAU{representation} of the Schr\"odinger \BLAU{Lie algebra}. 
Notably the two-point response function $R=\langle\phi\wit{\phi}\rangle$
is fixed up to an overall normalisation and (\ref{2reponse}) represents the complete information required, if we only add the second relation (\ref{gl:Cphaseord}), 
specific to phase-ordering kinetics. 
{}From {the hypothesis of the Schr\"odinger-covariance of $\langle\phi\phi\wit{\phi}\wit{\phi}\rangle$,} we can then derive, resp. re-derive, the following:
\begin{enumerate}
\item the algebraic long-time asymptotics (\ref{gl:3a}) of the two-time correlator, see also figure~\ref{fig1}d.
\item the exponent equality (\ref{lambda}) between auto-correlation and auto-response exponents.

The non-triviality of $\lambda$ is illustrated by a new scaling relation (\ref{gl:3.6}) with
a passage exponent. Eq.~(\ref{gl:3.6}) also reconfirms the long-standing bounds $d/2\leq\lambda\leq d$ \cite{Fish88a,Yeun96a} on the autocorrelation exponent.
\item the scaling (\ref{gl:2}) of the single-time correlator. We also showed that the resulting scaling forms are fully compatible with {\sc Porod}'s law.
\BLAU{The addition of this feature constitutes the main advance, due to a better conceptual basis.} 
\item the finite-size scaling (\ref{gl:3.15}) of the plateau height ${C}_{\infty}^{(2)}$ of the two-time auto-correlator.

Several tests of this new scaling law are reviewed in \cite{Henk25c} and more are forthcoming.
\item the scaling form (\ref{gl:3.16}) of the global two-time correlator, including the extension (\ref{gl:3.17}),
to all temperatures $T<T_c$, of the well-known {\sc jss} {critical-point} scaling relation \cite{Jans89}.
\item this also implies the scaling (\ref{gl:3.18}) of the global squared order-parameter and the finite-size scaling (\ref{gl:3.20})
of the plateau height  $\wht{C}_{\infty}^{(2)}$ in a fully finite system.
\end{enumerate}
All these results are either well-known {\it folklore} in phase-ordering kinetics \cite{Bray94a} or else new predictions which call for further tests in specific models. 
We have reviewed recently existing confirmations in detail \cite{Henk25c}. 
Our present approach has the \BLAU{clear} conceptual advantage that it only uses the covariance of response functions as an input. 

\BLAU{In addition, restricting ourselves to the covariance of merely the responses, offers the possibility of generalisation. This is welcome, since}  
at present, the use of the Schr\"odinger algebra $\mathfrak{sch}(1)$ does not provide enough constraints on four-point responses 
that statements on the \BLAU{explicit} form of $C(s;r)$ or $C(t,s)$ could be made.
Possible extensions might include the following directions:
\begin{enumerate}
\item consider extensions $\mathfrak{sch}(1)\subset\mathfrak{conf}(3)$ which can be done in such a way that non-trivial dynamical symmetries of the
Schr\"odinger equations are obtained \cite{Henk03a,Henk04b,Lorenz07a}. Previous attempts in this direction were biased since unrealistic initial conditions had been used.
\item consider extensions $\mathfrak{sch}(1)\subset\mathfrak{sv}(1)$ towards the infinite-dimensional Schr\"odinger-Virasoro algebra \cite{Henk94,Unte12}
and follow the programme analogous to the one carried out in \cite{BPZ} in $2D$ conformal field-theory.
It is possible that advances in non-relativistic holography \cite{Beka12,Fuer09,Gray13,Golk14,Shim21,Son08,Volo09} or operator-product-expansions \cite{Shim21,Paga25} could be of use.
Alternatively, could techniques borrowed from the $3D$ conformal bootstrap \cite{Pola19,Rych23} provide further insight~?
\item is there any conceptual input to be drawn from non-trivial solvable models such as the $(1+1)D$ Calogero model \cite{Shim21}~?
\end{enumerate}
\BLAU{The decisive test of Schr\"odinger-invariance in phase-ordering kinetics will be the explicit derivation of the correlator scaling functions $F_C$ in (\ref{gl:2}) 
and their comparison with results from specific models.} 
Work on these directions is in progress and the three lines mentioned here are in no way mutually exclusive.
Our study assumed the order-parameter $\phi$ (the magnetisation) to be quasi-primary with respect to the Schr\"odinger algebra $\mathfrak{sch}(1)$. The techniques presented could be
rapidly extended to energy-energy correlators $\langle \epsilon\epsilon\rangle$, provided the energy-density scaling operator $\epsilon$
(and also its response operator $\wit{\epsilon}$\,) is quasi-primary as well, see \cite{Baum07} for an exploration. We treated here classical systems throughout; investigations on
possible extensions to quantum dynamics \cite{Giam16} are highly desirable.

Our identification of the non-equilibrium representation (\ref{gl:4.1}) and of the rules how to use it, is empirical, 
although there is much evidence that it should be the correct choice in many cases.
We still lack any argument why this should be so.

Finally,  any attempt to get beyond the present restriction to $\mathpzc{z}=2$ would require to find a convincing candidate to describe time-space responses which could also distinguish
between non-conserved and conserved dynamics.

\noindent
{\bf Acknowledgements:}
We are grateful to D. Karevski, N. Stoilova and V.K. Dobrev for useful discussions.
This work was supported by the french ANR-PRME UNIOPEN (ANR-22-CE30-0004-01) and by PHC RILA (Dossier 51305UC/KP06-Rila/7).
\newpage

\appsection{A}{Four-point response function}

We compute the form of the four-point function
\BEQ
V=V\bigl(t,s,\vep,\vep';\vec{r},\vec{r}',\vec{R},\vec{R}'\bigr)=\left\langle \phi(t,\vec{r})\phi(s,\vec{r}')\wit{\phi}(\vep,\vec{R})\wit{\phi}(\vep',\vec{R}')\right\rangle_0
\EEQ
as a deterministic average at equilibrium.\footnote{The notation is from the german {\it Vier}punktfunktion.}
The initial times $\vep,\vep'$ are assumed to be much smaller than the observation/waiting times $t,s$ and will be specified in (\ref{gl:limitA9}) below.

We must work out the consequences of covariance under the generators of the Schr\"odinger algebra. In our notation, we either refer to the $1D$ case or consider the
spatial translations only in the radial direction such that $\vec{r}\cdot\frac{\partial}{\partial\vec{r}}=r\partial_r$ so that we need not study spatial rotations explicitly.
A quasi-primary scaling operator $\phi$ is characterised by the exponent $\delta$ and the mass $\cal M$.  From the two-point function (\ref{2reponseR}) we know that $\wit{\delta}={\delta}$ and
$\wit{\cal M}=-{\cal M}$. By construction, the mass conservation under $M_0$ is satisfied. Time-translation-invariance (generator $X_{-1}$)
is taken into account by working with the
reduced variables $\tau = t-s$, $\rho = \vep - s$ and $\rho' = \vep' - s$. Analogously, spatial translation-invariance (generator $Y_{-\demi}$) leads to
\BEQ
V = V\bigl(\tau,\rho,\rho';\vec{r}-\vec{r}',\vec{R}-\vec{r}',\vec{R}'-\vec{r}'\bigr)
\EEQ
Next, using spatial translation-invariance in the generator $Y_{\demi}$ of Galilei-transformations leads to
\BEQ \label{gl:A3}
Y_{\demi}V = \left[ -\tau\frac{\partial}{\partial r} - \rho \frac{\partial}{\partial R} - \rho'\frac{\partial}{\partial R'}
-{\cal M}\bigl( r-r'\bigr) +{\cal M}\bigl(R-r'\bigr) +{\cal M}\bigl( R'-r'\bigr) \right] V =0
\EEQ
Similarly, using both temporal and spatial translation-invariance in the dilatation generator $X_0$ gives
\BEQ \label{gl:A4}
X_0 V = \left[ -\tau\frac{\partial}{\partial\tau} -\rho\frac{\partial}{\partial \rho} - \rho'\frac{\partial}{\partial \rho'}
-\demi \bigl(r-r'\bigr)\frac{\partial}{\partial r} -\demi \bigl(R-r'\bigr)\frac{\partial}{\partial R} -\demi \bigl(R'-r'\bigr)\frac{\partial}{\partial R'} -4\delta \right] V =0
\EEQ
and where we used $\partial_r,\partial_R,\partial_{R'}$ as short-hands for the derivatives with respect to $r-r'$, $R-r'$ and $R'-r'$.
Finally, using the translation-invariances in the generator $X_1$ of special Schr\"odinger-invariance gives the covariance condition
{\small \BEA
\lefteqn{ X_1 V = \left[ -\bigl( t^2 - s^2\bigr)\frac{\partial}{\partial\tau} - \bigl(\vep^2 - s^2\bigr)\frac{\partial}{\partial\rho} -\bigl(\vep'^2-s^2\bigr)\frac{\partial}{\partial\rho'}
-\bigl(tr-sr'\bigr)\frac{\partial}{\partial r} -\bigl(\vep R-s r'\bigr)\frac{\partial}{\partial R} -\bigl(\vep' R-s r'\bigr)\frac{\partial}{\partial R'} \right.}
\nonumber \\
& & \left. -2\delta \tau - 2\delta\vep - 2 \delta\vep' - 8\delta s
-\frac{\cal M}{2}\bigl(r^2 - r'^2\bigr) +\frac{\cal M}{2}\bigl( R^2 -r'^2\bigr) +\frac{\cal M}{2}\bigl(R'^2-r'^2\bigr) \right] V
\\
&=& ~~\left[ -\tau^2\frac{\partial}{\partial\tau} -\rho^2\frac{\partial}{\partial\rho} -\rho'^2\frac{\partial}{\partial\rho'}
-\bigl(\tau (r-r')\bigr)\frac{\partial}{\partial r} -\bigl(\rho (R-r')\bigr)\frac{\partial}{\partial R} -\bigl(\rho' (R'-r')\bigr)\frac{\partial}{\partial R'} \right.
\nonumber \\
&& \left. ~~~-2\delta\tau -2\delta\rho - 2\delta\rho'
-\frac{\cal M}{2}\bigl(r- r'\bigr)^2 +\frac{\cal M}{2}\bigl( R -r'\bigr)^2 +\frac{\cal M}{2}\bigl(R'-r'\bigr)^2 \right] V
\nonumber \\
&& +2s\left[ -\tau\frac{\partial}{\partial\tau}-\rho\frac{\partial}{\partial\rho}-\rho'\frac{\partial}{\partial\rho'}
-\demi \bigl(r-r'\bigr)\frac{\partial}{\partial r} -\demi \bigl(R-r'\bigr)\frac{\partial}{\partial R} -\demi \bigl(R'-r'\bigr)\frac{\partial}{\partial R'} -4\delta\right] V
\nonumber \\
&& +r' \left[ -\tau\frac{\partial}{\partial r} - \rho \frac{\partial}{\partial R} - \rho'\frac{\partial}{\partial R'}
-{\cal M}\bigl( r-r'\bigr) +{\cal M}\bigl(R-r'\bigr) +{\cal M}\bigl( R'-r'\bigr) \right] V =0
\label{gl:A6}
\EEA}
Herein, the last two lines in (\ref{gl:A6}) automatically vanish because of Galilei- and dilatation-invariance (\ref{gl:A3},\ref{gl:A4}). In what follows, we perform a spatial
translation such that $r-r'\mapsto r$, $R-r'\mapsto R$ and $R'-r'\mapsto R'$. Further simplification is achieved by the changes of variables
\BEQ \label{gl:changeA7}
u = \demi\bigl( \rho + \rho'\bigr) \;\;,\;\; v = \demi\bigl( \rho - \rho'\bigr) ~~\mbox{\rm and}~~
\mathfrak{R} = \demi\bigl( R + R'\bigr) \;\;,\;\; \lap{\mathfrak{R}} = \demi\bigl( R - R'\bigr)
\EEQ
The first two of these changes produce from (\ref{gl:A3},\ref{gl:A4},\ref{gl:A6}) the system
\BEA
&& \left[ \tau\frac{\partial}{\partial r} -\bigl(u+v\bigr)\frac{\partial}{\partial R} - \bigl(u-v\bigr)\frac{\partial}{\partial R'} -{\cal M}r +{\cal M}\bigl( R+R'\bigr)\right]V=0
\nonumber \\
&& \left[ -\tau\frac{\partial}{\partial\tau}-u\frac{\partial}{\partial u}-v\frac{\partial}{\partial v}
-\demi r\frac{\partial}{\partial r} -\demi R\frac{\partial}{\partial R} -\demi R' \frac{\partial}{\partial R'} -4\delta\right] V =0
\label{gl:A8} \\
&& \left[ \tau^2\frac{\partial}{\partial\tau} -\bigl(u^2+v^2\bigr)\frac{\partial}{\partial u} -2uv\frac{\partial}{\partial v}
-\tau r\frac{\partial}{\partial}{\partial r} -\big(u+v\bigr)R\frac{\partial}{\partial R} -\bigl(u-v\bigr)R'\frac{\partial}{\partial R'} \right.
\nonumber \\
&& \left. ~~-2\delta\tau -4\delta u -\frac{\cal M}{2} r^2 + \frac{\cal M}{2} \bigl(R^2 + R'^2\bigr) \right]V=0
\nonumber
\EEA
Since both times $\vep,\vep'$ refer to the initial correlations, they will be small. We then can take the limits
\BEQ \label{gl:limitA9}
v = \demi\bigl( \vep - s -\vep' + s\bigr) = \demi \bigl( \vep - \vep'\bigr) \to 0 \;\; ,\;\;
u = \demi\bigl( \vep - s +\vep' - s\bigr) = \frac{\vep+\vep'}{2} -s         \to -s
\EEQ
Together with the last two changes of variables in (\ref{gl:changeA7}) this produces the system
\BEA
&&\left[ -\tau\frac{\partial}{\partial r} + s\frac{\partial}{\partial\fR} -{\cal M}r +2{\cal M}\,\fR\right]V=0
\nonumber \\
&&\left[ -\tau\frac{\partial}{\partial\tau}-s\frac{\partial}{\partial s} -\demi r\frac{\partial}{\partial r} -\demi \fR\frac{\partial}{\partial\fR}
-\demi\bR\frac{\partial}{\partial\bR} -4\delta \right] V = 0
\label{gl:A10} \\
&&\left[-\tau^2\frac{\partial}{\partial\tau} +s^2\frac{\partial}{\partial s}-2r\frac{\partial}{\partial r} +s\fR\frac{\partial}{\partial\fR}+s\bR\frac{\partial}{\partial\bR}
-2\delta\tau +4\delta s -\frac{\cal M}{2} r^2 +{\cal M}\bigl( \fR^2 + \bR^2\bigr)\right] V=0
\nonumber
\EEA
and where now $V=V\bigl(\tau,s;r,\fR,\bR\bigr)$. This will be basis for our further analysis where we must distinguish two cases.

\subsection{Two distinct times with $\tau>0$}

For non-vanishing $\tau$, the mass-dependent terms in (\ref{gl:A10}) are best parametrised by a further change of variable
\BEQ
V = \exp\left(-\frac{\cal M}{2} \frac{r^2}{\tau}\right) \exp\left( -{\cal M}\frac{\fR^2}{s} \right) \exp\left( -{\cal M}\frac{\bR^2}{s}\right)
G\bigl(\tau,s;r,\fR,\bR\bigr)
\EEQ
such that we have the more simple system
\BEA
&&\left[ -\tau\frac{\partial}{\partial r} + s\frac{\partial}{\partial\fR} \right] G = 0
\nonumber \\
&&\left[ -\tau\frac{\partial}{\partial\tau}-s\frac{\partial}{\partial s} -\demi r\frac{\partial}{\partial r} -\demi \fR\frac{\partial}{\partial\fR}
-\demi\bR\frac{\partial}{\partial\bR} -4\delta \right] G = 0
\label{gl:A12} \\
&& \left[-\tau^2\frac{\partial}{\partial\tau} +s^2\frac{\partial}{\partial s}-2r\frac{\partial}{\partial r} +s\fR\frac{\partial}{\partial\fR}+s\bR\frac{\partial}{\partial\bR}
-2\delta\tau +4\delta s \right] G=0
\nonumber
\EEA
In order to absorb the terms with the scaling dimension $\delta$, we now consider
\BEQ
G = \tau^{\gamma_1} s^{\gamma_2} (\tau+s\bigr)^{\gamma_3} H
\EEQ
Insertion into (\ref{gl:A12}) leads to the linear system of equations
\BEQ
\gamma_1 + \gamma_2 + \gamma_3 = -4\delta \;\; , \;\;
\gamma_1 + \gamma_3 = - 2\delta \;\; , \;\;
\gamma_2 + \gamma_3 = -4\delta
\EEQ
with the unique solution $\gamma_1=0$, $\gamma_2=\gamma_3=-2\delta$. Our final equations for the four-point response
\BEQ \label{gl:A15}
V = \exp\left[{-\frac{\cal M}{2}\frac{r^2}{\tau} -{\cal M}\frac{\fR^2}{s} -{\cal M}\frac{\bR^2}{s}}\right] \,s^{-2\delta} (\tau+s)^{-2\delta}\, H \;\; , \;\;
H = H\bigl(\tau,s;r,\fR,\bR\bigr)
\EEQ
are
\BEA
&&\left[ -\tau\frac{\partial}{\partial r} + s\frac{\partial}{\partial\fR} \right] H = 0
\nonumber \\
&&\left[ -\tau\frac{\partial}{\partial\tau}-s\frac{\partial}{\partial s} -\demi r\frac{\partial}{\partial r} -\demi \fR\frac{\partial}{\partial\fR}
-\demi\bR\frac{\partial}{\partial\bR}  \right] H = 0
\label{gl:A16} \\
&& \left[-\tau^2\frac{\partial}{\partial\tau} +s^2\frac{\partial}{\partial s}-2r\frac{\partial}{\partial r}
+s\,\fR\frac{\partial}{\partial\fR}+s\,\bR\frac{\partial}{\partial\bR} \right] H=0
\nonumber
\EEA
Standard techniques \cite{Kamk79} applied to (\ref{gl:A16}) now give the general solution in (\ref{gl:A15})
\BEQ
H = \mathscr{F}^{(2)}\left( \frac{\fR\,\tau + r\,s}{\bigl[\tau\,s(\tau+s)\bigr]^{1/2}}, \frac{\bR\, \tau^{1/2}}{\bigl[s (\tau+s)\bigr]^{1/2}} \right)
\;\; ; \;\; \mbox{\rm ~~if~~} \tau\ne 0
\EEQ
where $\mathscr{F}^{(2)}$ remains an unknown (differentiable) function of two variables. 
For $t=\tau+s \gg s$ this is (\ref{eq:2.8}) in the main text, generalised to any dimension $d>1$.

\subsection{Single time with $\tau=0$}

The equal-time case with $\tau=0$ cannot be obtained as a regular limit from the previous discussion.
Rather, we integrate the first condition of (\ref{gl:A10}) directly. Setting
\BEQ
V = \exp\left( -{\cal M}\frac{\fR^2}{s} \right) \exp\left( {\cal M}\frac{r\, \fR}{s}\right) \exp\left( -{\cal M}\frac{\bR^2}{s}\right)
G\bigl(\tau,s;r,\bR\bigr)
\EEQ
the remaining two conditions (\ref{gl:A10}) reduce to
\BEA
&& \left[ -s\frac{\partial}{\partial s} -\demi r\frac{\partial}{\partial r} -\demi \bR\frac{\partial}{\partial\bR} -4\delta\right] G =0
\nonumber \\
&& \left[ s^2\frac{\partial}{\partial s} +s\,\bR\frac{\partial}{\partial\bR} +4\delta s\right] G = 0
\label{gl:A19}
\EEA
Dividing the second of these by $s$ and then adding to the first one leads to
\BEQ
\left[ -r \frac{\partial}{\partial r} + \bR\frac{\partial}{\partial \bR} \right] G = 0
\EEQ
which is solved by $G=H\bigl(s,r\,\bR\bigr)$. Let $\mathfrak{u}=r\,\bR$. Substituted into the last condition  (\ref{gl:A19}) leads to the differential equation
\BEQ
\left[ s\frac{\partial H}{\partial s} + \mathfrak{u}\frac{\partial H}{\partial\mathfrak{u}} + 4\delta \right] H=0
\EEQ
such that finally
\BEQ
V = \exp\left[{-{\cal M}\frac{\fR^2}{s} + {\cal M}\frac{r\, \fR}{s} -{\cal M}\frac{\bR^2}{s}}\right] s^{-4\delta}\,
\mathscr{F}^{(1)}\left( \frac{r\,\bR}{s}\right) \;\; ; \;\; \mbox{\rm ~~if~~} \tau=0
\EEQ
where $\mathscr{F}^{(1)}$ is an unknown (differentiable) function of a single variable. This is (\ref{eq:2.9}) in the main text, generalised to $d>1$. 

\newpage
\appsection{B}{Finite-size four-point responses}

We generalise the analysis of the four-point function $V$ to the case of a finite system, of size $N$. In this case, only $\tau\ne 0$ is required.
We follow the habit of finite-size scaling theory, in and out of equilibrium \cite{Fish71,Barb83,Suzuki77}, and consider the relevant variable $\frac{1}{N}$.
Using the translation-invariances as in appendix~A, we have $V=V\bigl(\tau,\rho,\rho';r,R,R';\frac{1}{N}\bigr)$ where an implicit translation by $r'$ was performed.
The covariance conditions of Schr\"odinger-invariance are extended to
\BEA
&& \left[ -\tau\frac{\partial}{\partial r} -\rho\frac{\partial}{\partial R} -\rho'\frac{\partial}{\partial R'} -{\cal M}r +{\cal M}\bigl(R+R'\bigr) \right] V = 0
\nonumber \\
&& \left[ -\tau\frac{\partial}{\partial\tau} -\rho\frac{\partial}{\partial\rho}-\rho'\frac{\partial}{\partial\rho'}
-\demi r\frac{\partial}{\partial r} -\demi R\frac{\partial}{\partial R} -\demi R'\frac{\partial}{\partial R'}
-4\delta +\demi \frac{1}{N}\frac{\partial}{\partial \frac{1}{N}} \right]V=0
\\
&&\left[-\tau^2\frac{\partial}{\partial \tau} -\rho^2\frac{\partial}{\partial \rho} -\rho'^2\frac{\partial}{\partial \rho'}
-\tau\, r\frac{\partial}{\partial r} -\rho\, R\frac{\partial}{\partial R} -\rho'\, R'\frac{\partial}{\partial R'} \right.
\nonumber \\
&& \left.~~-2\delta\tau -2\delta\rho -2\delta\rho' -\frac{\cal M}{2}r^2 +\frac{\cal M}{2}\bigl( R^2 +R'^2\bigr)
+ \demi \bigl(\tau+\rho+\rho'\bigr) \frac{1}{N}\frac{\partial}{\partial \frac{1}{N}} \right]V=0
\nonumber
\EEA
where the extra term describing the finite-size effects are adapted from the two-point function considered in \cite{Henk25c} to the four-point function at hand.
In particular, we took into account that both $\vep,\vep' \ll t,s$ and checked the consistency with the commutators of the Schr\"odinger Lie algebra.  Adopting once more the
changes of variables (\ref{gl:changeA7}) and then the limit (\ref{gl:limitA9}), the conditions (\ref{gl:A10}) are extended into
\BEA
&&\left[ -\tau\frac{\partial}{\partial r} + s\frac{\partial}{\partial\fR} -{\cal M}r +2{\cal M}\,\fR\right]V=0
\nonumber \\
&&\left[ -\tau\frac{\partial}{\partial\tau}-s\frac{\partial}{\partial s} -\demi r\frac{\partial}{\partial r} -\demi \fR\frac{\partial}{\partial\fR}
-\demi\bR\frac{\partial}{\partial\bR} -4\delta +\demi \frac{1}{N}\frac{\partial}{\partial \frac{1}{N}}\right] V = 0
\label{gl:B2} \\
&&\left[-\tau^2\frac{\partial}{\partial\tau} +s^2\frac{\partial}{\partial s}-2r\frac{\partial}{\partial r} +s\fR\frac{\partial}{\partial\fR}+s\bR\frac{\partial}{\partial\bR}  \right.
\nonumber \\
&& \left. ~~-2\delta\tau +4\delta s -\frac{\cal M}{2} r^2 +{\cal M}\bigl( \fR^2 + \bR^2\bigr) +\demi \bigl(\tau-2s\bigr) \frac{1}{N}\frac{\partial}{\partial \frac{1}{N}}\right] V=0
\nonumber
\EEA
As in appendix~A, we can make the change
\BEQ \label{gl:B3}
V = e^{-\frac{\cal M}{2}\frac{r^2}{\tau} -{\cal M}\frac{\fR^2+\bR^2}{s}}\,s^{-2\delta} (\tau+s)^{-2\delta}\, H \;\; , \;\;
H = H\bigl(\tau,s;r,\fR,\bR;\frac{1}{N}\bigr)
\EEQ
which simplifies (\ref{gl:B2}) into
\BEA
&&\left[ -\tau\frac{\partial}{\partial r} + s\frac{\partial}{\partial\fR} \right] H = 0
\nonumber \\
&&\left[ -\tau\frac{\partial}{\partial\tau}-s\frac{\partial}{\partial s} -\demi r\frac{\partial}{\partial r} -\demi \fR\frac{\partial}{\partial\fR}
-\demi\bR\frac{\partial}{\partial\bR} +\demi \frac{1}{N}\frac{\partial}{\partial \frac{1}{N}} \right] H = 0
\label{gl:B4} \\
&& \left[-\tau^2\frac{\partial}{\partial\tau} +s^2\frac{\partial}{\partial s}
-2r\frac{\partial}{\partial r} +s\,\fR\frac{\partial}{\partial\fR}+s\,\bR\frac{\partial}{\partial\bR} +\demi \bigl(\tau-2s\bigr)
\frac{1}{N}\frac{\partial}{\partial \frac{1}{N}} \right] H=0
\nonumber
\EEA
whose standard solution \cite{Kamk79}, with the unknown function $\mathscr{F}^{(2,N)}$, is
\BEQ \label{gl:B5}
H = \mathscr{F}^{(2,N)}\left(
\frac{\fR\,\tau + r\,s}{\bigl[\tau\,s(\tau+s)\bigr]^{1/2}}, \frac{\bR\, \tau^{1/2}}{\bigl[s (\tau+s)\bigr]^{1/2}}, \frac{1}{N} \frac{\tau+s}{\tau^{1/2}} \right)
\EEQ

\newpage
\appsection{C}{Two-time response function}

We reconsider the calculation of the form of the two-time auto-response function $R=\bigl\langle\phi\wit{\phi}\bigr\rangle$, in the presence of a further dimensionful parameter.
To be definite, we shall use the case of an initially magnetised state, but the same discussion could be carried out for finite-size effects or a dimensionful coupling constant.
It is enough to restrict to the case of Schr\"odinger-invariance,
since the extension to non-equilibrium situations is by now well-known \cite{Henk25c}, but $\mathpzc{z}$ can be maintained
arbitrary, since we only deal with auto-responses.

First, we must write the Lie algebra generators. The ones of time-translations $X_{-1}$ and dilatations $X_0$ are
\begin{subequations}
\begin{align}
X_{-1} &= -\partial_t - \partial_s \\
X_0    &= -t\partial_t - s\partial_s - \frac{x_0}{\mathpzc{z}} m_0 \partial_{m_0} -2\delta + \frac{d}{\mathpzc{z}}
\end{align}
where $m_0$ is the initial magnetisation, $\frac{x_0}{\mathpzc{z}}$ describes its dimension and we took into account that we require the global auto-response
$\wht{R}(t,s;\vec{0};m_0)$ \cite{Henk25,Henk25c}.
We complete this by making the following ansatz for the generator $X_1$ of special transformations
\begin{align} \label{eq:C1c}
X_1 &= -t^2\partial_t - s^2\partial_t -\mu(t,s,m_0) \frac{x_0}{\mathpzc{z}} m_0 \partial_{m_0} -\alpha(t,s,m_0)
\end{align}
\end{subequations}
and where $\mu,\alpha$ must be fixed such that one retrieves the commutators $[X_1,X_{-1}]=2X_0$ and $[X_1,X_0]=X_1$. The first of these produces the conditions
\begin{subequations}
\BEQ
\demi\left( \frac{\partial\mu}{\partial t} + \frac{\partial \mu}{\partial s}\right) = 1 \;\; , \;\;
\demi\left( \frac{\partial\alpha}{\partial t} + \frac{\partial \alpha}{\partial s}\right) = 2\delta-\frac{d}{\mathpzc{z}}
\EEQ
whereas the second one gives the further conditions
\BEQ
t\frac{\partial\mu}{\partial t} +s\frac{\partial\mu}{\partial s} +\frac{x_0}{\mathpzc{z}} m_0\frac{\partial\mu}{\partial m_0} = \mu \;\; , \;\;
t\frac{\partial\alpha}{\partial t} +s\frac{\partial\alpha}{\partial s} +\frac{x_0}{\mathpzc{z}} m_0\frac{\partial\alpha}{\partial m_0} = \alpha
\EEQ
\end{subequations}
It follows that both $\mu$ and $\alpha$ must be linear functions in $t$ and $s$. If we write
\BEQ
\mu = \mu_1(m_0)t + \mu_2(m_0) s +\mu_3(m_0) \;\;,\;\;
\alpha= \alpha_1(m_0)t + \alpha_2(m_0) s +\alpha_3(m_0)
\EEQ
then it follows
\BEQ
\mu_1 + \mu_2 = 2 \;\; , \;\; \alpha_1 + \alpha_2 = 2\bigl(2\delta -\frac{d}{\mathpzc{z}}\bigr)
\EEQ
and $\mu_3\sim m_0^{-\mathpzc{z}/x_0}$ and $\alpha_3\sim m_0^{1-\mathpzc{z}/x_0}$. Then the special Lie algebra generator is
\addtocounter{equation}{-4}
\begin{subequations}
\addtocounter{equation}{3}
\begin{align} \label{eq:C1d}
X_1 &= -t^2\partial_t -s^2\partial_s -\bigl(1+\mu\bigr)t\,m_0\partial_{m_0} -\bigl(1-\mu\bigr) s\,m_0\partial_{m_0} -\mu_0 m_0^{1-\mathpzc{z}/x_0}\partial_{m_0} \nonumber \\
    &~~~-\bigl(2\delta-\frac{d}{\mathpzc{z}}\bigr)\bigl(1+\alpha\bigr)t -\bigl(2\delta-\frac{d}{\mathpzc{z}}\bigr)\bigl(1-\alpha\bigr)s -\alpha_0m_0^{\mathpzc{z}/x_0}
\end{align}
\end{subequations}
\addtocounter{equation}{3}
\noindent
which gives the concrete form of (\ref{eq:C1c}). Herein, $\mu,\mu_0,\alpha,\alpha_0$ are free parameters which do not impact on the commutator relation.

We now inquire about the consequences of their presence for the global
two-time auto-response $\wht{R}_0=\wht{R}_0\bigl(t,s;\vec{0};m_0\bigr)$.
Time-translation-invariance simply leads to $\wht{R}_0=\wht{R}_0\bigl(\tau,m_0\bigr)$ with $\tau=t-s$.
Dilatation-invariance, by using time-translation-invariance as well, first gives the equation
\BEQ
\left[ -\tau\partial_{\tau}-\frac{x_0}{\mathpzc{z}}m_0\partial_{m_0} -\bigl(2\delta-\frac{d}{\mathpzc{z}}\bigr) \right] \wht{R}_0 =0
\EEQ
which is solved by \cite{Henk25c}
\BEQ \label{gl:C6}
\wht{R}_0 = \tau^{-2\delta+d/\mathpzc{z}} \mathscr{F}_m\left( m_0 \tau^{-x_0/\mathpzc{z}} \right)
\EEQ
and by physical consistency must obey the boundary conditions (i) $\mathscr{F}_m\bigl(0\bigr)=1$ and
(ii) $\mathscr{F}\bigl(\mathfrak{m}\bigr)\stackrel{\mathfrak{m}\gg 1}{\sim} \mathfrak{m}^{-1}$.
After the standard extension to non-equilibrium scaling \cite{Henk25c} and for $t\gg s$, this indeed reproduces the well-known scaling of the time-dependent magnetisation
$m(t)=m_0 \wht{R}_0$ \cite{Jans89,Taeu14} at criticality and with an initial value $m_0$.

Does the special Schr\"odinger-invariance furnish additional information on the scaling function $\mathscr{F}_m$~?
Combining time-translation-invariance with the explicit generator (\ref{eq:C1d}) we find
\BEA
\left[ -\tau^2\partial_{\tau} -\bigl(1+\mu\bigr)\frac{x_0}{\mathpzc{z}} \tau m_0 \partial_{m_0} -\bigl(2\delta-\frac{d}{\mathpzc{z}}\bigr)\bigl(1+\alpha\bigr)\tau
-\mu_0 m_0^{1-\mathpzc{z}/x_0} \partial_{m_0} -\alpha_0 m^{-\mathpzc{z}/x_0} \right] \wht{R}_0 =0
\EEA
Inserting the scaling from (\ref{gl:C6}) into this, we find
\BEQ
\tau\,\left[ -\alpha \bigl(2\delta-\frac{d}{\mathpzc{z}}\bigr)\mathscr{F}_m(\mathfrak{m}) -\mu \frac{x_0}{\mathpzc{z}} \mathfrak{m}\mathscr{F}_m'(\mathfrak{m}) \right]
+ m_0^{-\mathpzc{z}/x_0} \biggl[ -\mu_0 \mathfrak{m}\mathscr{F}_m'(\mathfrak{m}) -\alpha_0 \mathscr{F}_m(\mathfrak{m}) \biggr] = 0~~~~
\label{gl:C8}
\EEQ
The two pre-factors exclude a common dependence of both terms on the single scaling variable $\mathfrak{m}=m_0 \tau^{-x_0/\mathpzc{z}}$.
Then in at most one of the brackets in (\ref{gl:C8}) a compensation
of the terms inside can occur while the parameters in the other bracket must vanish. In both cases, this leads to a pure power-law
\BEQ \label{eq:C9}
\mathscr{F}_m(\mathfrak{m}) = \mathscr{F}_0\, \mathfrak{m}^{-f} \;\; , \;\; f = \frac{\alpha_0}{\mu_0} = \frac{2\delta\mathpzc{z}-d}{x_0}\cdot\frac{\alpha}{\mu}
\EEQ
While it would be possible to arrange for $f=1$, eq.~(\ref{eq:C9}) then is not consistent with the boundary condition $\mathscr{F}_m(0)=1$.
It only remains that $\mu=\mu_0=\alpha=\alpha_0=0$. This means that the scaling form (\ref{gl:C6}) is
also reproduced from the covariance $X_1 \wht{R}_0=0$ under the special transformations such that $\mathscr{F}_m$ is left un-determined.
Hence Schr\"odinger-invariance is not able to fix it and the question raised above about $\mathscr{F}_m$ is answered in the negative.

\newpage
\appsection{D}{On Porod's law}

The following presentation is based on \cite{Henk20}. In phase-ordering kinetics the {\em structure factor}
\BEQ
\wht{S}(t;\vec{q}) = \int_{\mathbb{R}^d} \!\D\vec{r}\: e^{-\II\vec{q}\cdot\vec{r}}\: C(t;\vec{r})
\EEQ
is defined as the Fourier transform of the single-time correlator $C(t;\vec{r})$. For a scalar order-parameter $\phi$, it was realised by {\sc Porod} long ago
that if $|\vec{q}|\gg 1$, asymptotically
$\wht{S}(t;\vec{q})\sim \bigl( \ell(t) |\vec{q}|^{d+1} \bigr)^{-1}$ \cite[eq. (42)]{Poro51},\cite{Cicc88,Gall12} where $\ell(t)\sim t^{1/2}$
is the growing length scale.\footnote{For generalisations to anisotropic or fractal samples see e.g. \cite{Bale84,Cicc00,Chia19}.}
For an $n$-component vector order-parameter one similarly suggests a {\em generalised Porod law} \cite{Bray92a,Bray94a}
\BEQ \label{gl:D2}
\wht{S}(t;\vec{q}) \stackrel{|\vec{q}|\gg 1}{\sim} \bigl( \ell(t)^n |\vec{q}|^{d+n} \bigr)^{-1}
\EEQ
The celebrate Wiener-Khintchine theorem \cite{Fell71} asserts, since the structure factor $\wht{S}$ is the Fourier transform of a spatially translation-invariant two-point correlator,
that $\wht{S}(t;\vec{q})\geq 0$ is proportional to a probability distribution. More precisely, the macroscopically observed structure factor $\wht{S}(t;\vec{q})$ should
be thought of as the sum of the scattering of many microscopic scatterers. Below the critical point, the spatial correlation length $\bar{\xi}(T)$ is finite such that microscopic
sites, more distant than $\bar{\xi}(T)$ from each other, should be independent. The central limit theorem \cite{Fell71} on the sum of independent scatterers states that for a large number
${\cal N}\gg 1$ the distribution of the sum should converge towards a gaussian, provided that the second moment of $\wht{S}(t,\vec{q})$ is finite. For a scalar order-parameter (hence $n=1$)
this second moment diverges, since with the use of (\ref{gl:D2})
\BEQ
\langle q^2 \rangle \sim \int_0^{\infty} \!\D q\: q^{d-1} q^2 \wht{S}(t;\vec{q}) \sim \int^{\infty} \!\D q\; \frac{q^{d+1}}{q^{d+1}} = \infty
\EEQ
but for a vector order-parameter with $n\geq 2$ one has, again with (\ref{gl:D2})
\BEQ
\langle q^2 \rangle \sim \int_0^{\infty} \!\D q\: q^{d-1}\, q^2\, \wht{S}(t;\vec{q}) \sim \int^{\infty} \!\D q\; \frac{q^{d+1}}{q^{d+n}} =
\left\{ \begin{array}{ll} \infty       & \mbox{\rm ~~;~ if $n=2$} \\
                          {\rm finite} & \mbox{\rm ~~;~ if $n>2$}
        \end{array} \right.
\EEQ
This suggests that for $n>2$, the central limit theorem should apply such that the form of $\wht{S}(t;\vec{q})$ for sufficiently many degrees of freedom will become gaussian;
whereas for $n=1$ the distribution of $\wht{S}(t;\vec{q})$ will be different and $n=2$ should be the marginal case between the two.
A very convincing experimental example of this occurs in the phase-ordering of the $3D$ binary alloy Cu$_3$Au \cite{Shan92}.\footnote{Since the sites of the ordered state
are occupied `antiferromagnetically' by the Cu and Au atoms, the order-parameter is the staggered magnetisation, which is non-conserved, in agreement with the
measured $\mathpzc{z}=2$.} For small times, the structure factor can indeed be successfully fitted by a gaussian distribution which for larger times crosses over to
a lorentzian-squared form \cite{Shan92}
\BEQ
\wht{S}(t;\vec{q}) \sim \left( \frac{\Gamma(t)^2}{\Gamma(t)^2 + \vec{q}^2}\right)^2 
~~\Longrightarrow~~ C(t;\vec{r}) = C_0\, e^{-\Gamma(t) |\vec{r}|} \;\; ; \;\; \Gamma(t)\sim t^{-0.50(3)} 
\EEQ
(see \cite[Ex.~1.10]{Henk10} for the calculation of $C(t;\vec{r})$) which does show the expected cusp at $|\vec{r}|=0$.

The gaussianity of the structure factor $\wht{S}(t;\vec{q})$ can be tested via the study of global persistence, see \cite{Bray13} for a thorough review.
In a finite spatial volume $\Omega$, consider the global order-parameter
$\wht{\phi}_0(t) := |\Omega|^{-1/2}\int_{\Omega} \!\D\vec{r}\: \phi(t,\vec{r})$
(and later send the volume $|\Omega|\to\infty$). From the usual dynamical scaling it follows that the global normalised correlator scales
\BEQ
\wht{N}(t_1,t_2) :=
\frac{ \bigl\langle \wht{\phi}_0(t_1) \wht{\phi}_0(t_2) \bigr\rangle}{\sqrt{\bigl\langle \wht{\phi}^2_0(t_1)\bigr\rangle\bigl\langle \wht{\phi}^2_0(t_2) \bigr\rangle\,}}
= f_{\wht{N}}\left( \frac{t_1}{t_2} \right)
\EEQ
The {\em global persistence probability} $P_{g}(t)$ is defined as the probability that a component of the global order-parameter $\wht{\phi}_0(t)$
did not change sign up to the time $t>0$ (and average over the various components if $n>1$). For long times, one expects the decay
$P_g(t) \sim t^{-\theta_g}$, where $\theta_g$ is the {\em global persistence exponent}. One has the following

\noindent
{\bf Theorem:} \cite{Cuei97a,Henk09a} {\it For a non-equilibrium spin system quenched to $T<T_c$ and which is such that
(i) the central limit theorem holds for $\wht{\phi}_0(t)$ (which is hence gaussian) and (ii) the dynamics of the spin variables is markovian, then}
\BEQ \label{persisglob}
\boxed{ \mbox{\rm (I)~~}  \theta_g = \frac{1}{\mathpzc{z}} \left( \lambda - \frac{d}{2}\right) \mbox{\rm ~~,~~~(II)~~}  f_{\wht{N}}(y) = y^{-\theta_g} \mbox{\rm ~~exactly~~}}
\EEQ

\begin{table}[tb]
\begin{tabular}{|l|clcl|ll|cc|} \hline
      &                          &            &                           &                                  & \multicolumn{2}{c|}{$\theta_g$} & & \\ \cline{6-9}
model & \multicolumn{1}{|c}{$d$} & condition~ & \multicolumn{1}{c}{~$z$~} & \multicolumn{1}{l|}{~~$\lambda$} & markovian~ & numeric~           & \multicolumn{2}{|c|}{~~Ref.~~} \\ \hline
Ising       & $1$  & $T=0$     & $2$ & $1$            & $1/4$          & $1/4$         & \cite{Godr00a}        & \cite{Maju96a} \\[0.09cm]
Ising       & $2$  & $T=0$     & $2$ & $1.24(2)$~     & $0.12(1)$      & $\simeq 0.09$ & \cite{Fish88a}        & \cite{Cuei97a}  \\
~           &      & $T=1.0$   & $2$ & $1.24(2)$~     & $0.12(1)$~     & $0.062(2)$~   &                       & \cite{Henk09a}  \\
~           &      & $T=1.5$   & $2$ & $1.24(2)$      & $0.12(1)$      & $0.065(2)$~   &                       & \cite{Henk09a}  \\[0.09cm]
{\sc tdgl}  & $2$  & $T=0$     & $2$ & $1.24$         & $0.12$         & $\simeq 0.06$ & \cite{Fish88a}        & \cite{Cuei98a}  \\[0.09cm]
XY          & $2$  & $T=0$     & $2$ & $\approx 1.25$ & $\approx 0.13$ & $0.22(1)$     & \cite{Abri03}         & \cite{Bhar10}  \\[0.09cm]
Heisenberg  & $3$  & $T=0$     & $2$ & $\approx 1.6$  & $\approx 0.1$  & $0.13(1)$     & \cite{Maze90,Bray92a} & \cite{Bhar10}  \\[0.09cm]
spherical~  & ~$>2$~ & $T<T_c$ & $2$ & $d/2$          & $0$            & $0$           & \cite{Godr00b}        & \cite{Henk09a}   \\[0.09cm]
long-range~ &      &           &     &                &                &               &                       & \\
spherical~  & ~$>\sigma$~ & $T<T_c$ & $\sigma$ & $d/2$ & $0$           & $0$           & \cite{Cann01}         & \cite{Henk09a}   \\ \hline
\end{tabular}
\caption[tab1]{Test of the markovian scaling relation (\ref{persisglob}) for the global persistence exponent
$\theta_g$ in phase-ordering kinetics for several models.
The long-ranged spherical model is considered for $0<\sigma<2$.
The quoted sources refer to the direct calculation of $\theta_g$, labelled `numeric'.
\label{tab1} }
\end{table}

Testing the scaling relation (\ref{persisglob}) provides a means to check for the gaussianity of the global order-parameter. In table~\ref{tab1} the results
of (\ref{persisglob}), referred to as `markovian', are compared with direct numerical simulations in several model quenched into their ordered phase.
The first column of references is on the values of $\lambda$ and the second column are direct numerical estimates of $\theta_g$ (which are $T$-independent \cite{Bhar10,Cuei97a,Henk09a}).
In the XY and Heisenberg models, the values of $\lambda$ are not known with a great precision such that a comparison with (\ref{persisglob}) is not easy.
But since in general one finds disagreement,
it follows that in general {\em the spin relaxation dynamics cannot be simultaneously markovian and have a gaussian distribution of $\wht{\phi}_0(t)$}.
Precise tests of this in the $3D$ vector model with O(3) or O(4)-symmetry would be of interest.

In carrying out such tests, since often the numerical estimates of $\theta_g$ are not very different from the markovian prediction (\ref{persisglob}), it
can be more conclusive to study the form of $f_{\wht{N}}(y)$, see \cite[fig.2]{Henk09a} for a clear example in the $2D$ Ising model.

\end{document}